\documentclass[aps,twocolumn,showpacs,superscriptaddress,amsmath,amssymb,prl,nofootinbib]{revtex4-1}
\usepackage{epsf,amsmath,amssymb,amsfonts,verbatim,color,multirow,pifont}
\usepackage{graphicx}
\usepackage{dcolumn}% Align table columns on decimal point
\usepackage{bm}% textbf math
\usepackage{txfonts}
\usepackage{hyperref}
\usepackage{soul}
\begin{document}

\title{Thermodynamic cost, speed, fluctuations, and error reduction of biological copy machines}

\author{Yonghyun Song}
\affiliation{Korea Institute for Advanced Study, Seoul 02455, Korea}
\author{Changbong Hyeon}
\affiliation{Korea Institute for Advanced Study, Seoul 02455, Korea}
\email{hyeoncb@kias.re.kr}

% Please give the surname of the lead author for the running footer

% Please include corresponding author, author contribution and author declaration information
%\authorcontributions{Author contributions: Y.H.S. and C.H. designed research; Y.H.S. and C.H. performed research; Y.H.S. and C.H. analyzed data; and Y.H.S. and C.H. wrote the paper.
%}
%\authordeclaration{The authors declare no competing interest.}
%\equalauthors{\textsuperscript{1}A.O.(Author One) and A.T. (Author Two) contributed equally to this work (remove if not applicable).}
%\correspondingauthor{\textsuperscript{1}To whom correspondence should be addressed. E-mail: hyeoncb@kias.re.kr}

% Keywords are not mandatory, but authors are strongly encouraged to provide them. If provided, please include two to five keywords, separated by the pipe symbol, e.g:

%\abbreviations{YS,CH}
%\keywords{kinetic proofreading $|$ kinetic discrimination $|$ thermodynamic uncertainty relation $|$ copy number fluctuations $|$ biological machines} 

\begin{abstract}
Due to large fluctuations in cellular environments, 
transfer of information in biological processes without regulation is inherently error-prone. 
The mechanistic details of error-reducing mechanisms in biological copying processes have been a subject of active research; however, how error reduction of a process is balanced with its thermodynamic cost and dynamical properties remain largely unexplored. 
Here, we study the error reducing strategies in light of the recently discovered thermodynamic uncertainty relation (TUR) that sets a physical bound to the cost-precision trade-off relevant in general dissipative processes.   
We found that the two representative copying processes, DNA replication by the exonuclease-deficient T7 DNA polymerase and mRNA translation by the \textit{E. coli} ribosome, reduce the  error rates to biologically acceptable levels while also optimizing the processes close to the physical limit dictated by TUR.
\end{abstract}
\maketitle

Biological copying processes, which include DNA replication, transcription, and translation, 
have evolved error-reducing mechanisms to faithfully transmit information in the genetic code. 
In their seminal papers in the 1970s, Hopfield and Ninio \cite{Hopfield1974,Ninio1975} proposed 
the \emph{kinetic proofreading mechanism} 
to show that the energy-burning action of the mechanism can reduce the error rate. Shortly after, Bennett showed that 
the difference between kinetic barriers involving the incorporation of correct and incorrect substrates could be capitalized on to reduce the error rate under nonequilibrium chemical driving forces \cite{Bennett1976}.
Despite differences in their mechanistic details, both models share a common feature that the reduction of copying error incurs free energy cost. 
Since these pioneering works, there have been a number of studies devoted to understanding the relation between the error reduction, speed, and energy consumption not only in the biological copying processes \cite{Banerjee2017,Cady2009,Mallory2019,Mellenius2017a,Murugan2012a,Rao2015}, but also in more general biochemical networks, including those related to sensory adaptation, circadian rhythm, and metabolic control \cite{Cao2015a, Hartich2015, Francois2016, Lan2013, Marsland2019, Qian2006}.  

Besides the faithful transmission of genetic information,
the primary goal of biological copying processes is to generate biomass in the forms of DNA, RNA, and proteins.
Intuitively, however, error reduction comes at the cost of energy dissipation or slowing down of the process. 
Furthermore, fluctuations in biomass synthesis, which concomitantly increase with heat dissipation for Michaelis-Menten type processes \cite{Hwang2017JPCL}, also have to be suppressed below a biologically acceptable level. 
For instance, DNA replication in early fly embryogenesis occurs at high speed with exquisite precision;  a modest change of 10 \% in replication timing could be lethal \cite{Djabrayan2019}.
Similarly, for translation, it is well known that cells must express genes at the right protein copy number for optimal function in a given environment \cite{Dekel2005,Scott2014,Li2014a}; 
regulatory mechanisms are developed to suppress the copy number fluctuation in gene expression \cite{Fraser2004}.
How biological processes balance these conflicting requirements is a fundamental subject to explore.  
To address such an issue, the recently developed \emph{thermodynamic uncertainty relation} (TUR) \cite{barato2015PRL}, which offers a quantitative bound for dissipative processes at nonequilibrium steady states (NESS), is well suited.

TUR expresses the trade-off between the thermodynamic cost and uncertainty of dynamical processes in NESS and specifies its physical bound as follows:
\begin{align}
\mathcal{Q} = q(t) \epsilon_X^2(t) \geq 2 k_BT. 
\label{eqn:TUR} 
\end{align}
This form of TUR holds for most of biological processes that can be represented either by stochastic jump processes on
a kinetic network or by overdamped Langevin dynamics \cite{Gingrich2016PRL,Hyeon2017PRE,dechant2018PRE,Pigolotti2017PRL}, though extensions to more general conditions, which adjust the lower bound of the original relation, have also been discussed in recent years \cite{lee2018PRE,horowitz2017PRE,brandner2018PRL,barato2018NJP,chun2019PRE,Marsland2019,hasegawa2019PRL,Timpanaro2019PRL,horowitz2019NaturePhys}. 
Briefly,
 $\epsilon_X(t)\equiv \sqrt{\langle \delta X(t)^2\rangle}/\langle X(t)\rangle$ is a relative uncertainty (or error) in an output observable $X(t)$ best representing the dynamic process at time $t$, and 
$q(t)$ denotes the thermodynamic cost or heat dissipation in generating the dynamic trajectory. 
The inequality in Eq.\ref{eqn:TUR} allows one to quantitatively assess the physical limit to the precision that a dynamical process can maximize for a given amount of dissipation. 
{\color{black}
Recently, $2k_BT/\mathcal{Q}$, which is bounded between 0 and 1, was used to quantify the ``transport efficiency'' of molecular motors \cite{dechant2018PRE}. 
When $\mathcal{Q}$ is written in the form of $\mathcal{Q}=\dot{q}(t)(2D/V^2)$ with $D$ and $V$ being the diffusivity and velocity of a molecular motor,  the motor characterized with a small $\mathcal{Q}$ can be interpreted as an efficient cargo transporter, because it transports cargos with high velocity ($V\sim \langle X(t)\rangle /t$), small fluctuation ($D\sim \langle \delta X(t)^2\rangle/2t$), but with small dissipation rate ($\dot{q}$) \cite{Hwang2018JPCL}. 
Biosynthetic reactions that are efficient in suppressing fluctuations in product formation can also be characterized by small $\mathcal{Q}$.
}

This work is organized into four parts. 
(i) We first introduce the basics of biological copying processes by reviewing the two distinct error reducing strategies by Bennett  \cite{Bennett1976} and Hopfield \cite{Hopfield1974}. 
(ii) We evaluate the error rate and $\mathcal{Q}$ of the replication process by the exonuclease-deficient T7 DNA polymerase, a model process reminiscent of the kinetic discrimination mechanism by Bennett.
(iii) We analyze a model of mRNA translation where both Bennett's kinetic discrimination and Hopfield's kinetic proofreading are employed to lower the error rate, and calculate $\mathcal{Q}$ for translating a codon into a polypeptide chain. 
(vi) Lastly, we consider a more realistic model of mRNA translation that explicitly accounts for 42 types of aa-tRNA, and show that kinetic proofreading can suppress the fluctuation in the rate of polypeptide production.

\section{Results} 
\begin{figure}[h]
	\includegraphics[width=0.475\textwidth]{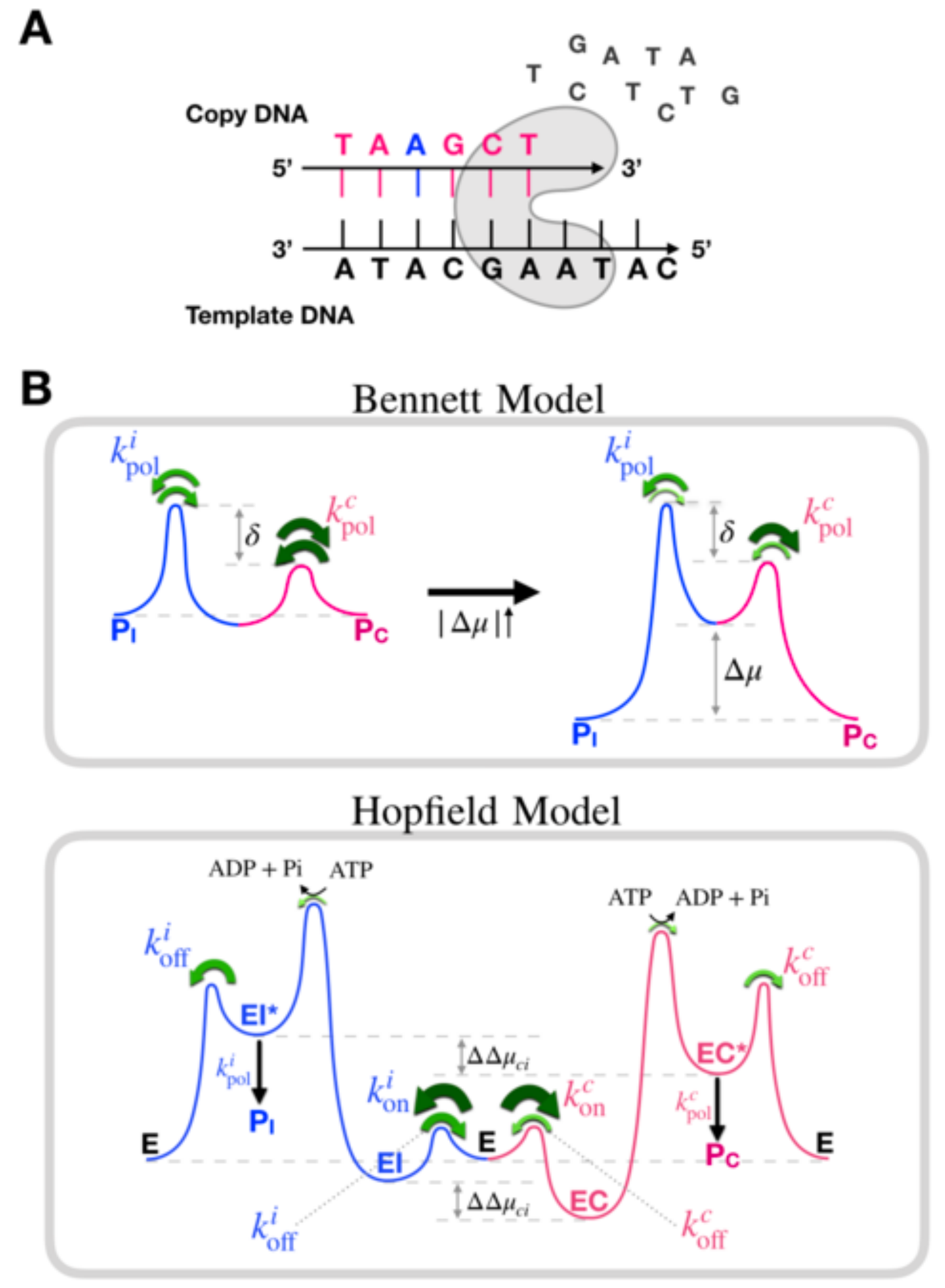}
	\caption{The error reducing mechanisms of Bennett and Hopfield.
	(A) A cartoon illustrating a biological copying process using DNA replication. When the sequence of template DNA is copied, complementary (correct, $c$) or non-complementary (incorrect, $i$) nucleotide can be incorporated into the copy DNA. 
	(B) (Top) The schematic of Bennett model \cite{Bennett1976}.
	Substrates are polymerized through a one-step enzyme reaction.
	Energetic input in the form of the chemical potential of the substrates ($\Delta\mu$) leads to a larger current of correct substrate incorporation and reduces the error probability.
	(Bottom) The schematic of Hopfield model \cite{Hopfield1974}.
	The substrates are polymerized through a three-state kinetic mechanism with intermediate states ${\bf E}$, ${\bf EC}$, and ${\bf EC}^*$ for correct substrate, or ${\bf E}$, ${\bf EI}$, and ${\bf EI}^*$ for incorrect substrate. 
	The reactions ${\bf EC}\rightarrow{\bf EC}^*$ and ${\bf EI}\rightarrow{\bf EI}^*$, which expend the chemical energy of ATP hydrolysis, are effectively irreversible. 
	This allows the copy process to select against the incorrect substrate through two chances of facilitated unbinding, thereby reducing the error probability. 
	For both schematics, the thickness and color of the arrows represent the relative magnitude of the respective rate constants.
	 }
	\label{fig:Schematics}
%	\end{center}
\end{figure}

{\bf Error reducing mechanisms by Bennett and Hopfield. }
We briefly describe the two representative error reducing mechanisms, one by Bennett and the other by Hopfield. 
In a nutshell, the essence of the two mechanisms lies in an energy-dissipating enzymatic reaction of copy machines comprised of multiple kinetic cycles that can discriminate correct substrates from incorrect ones. 
Illustrated in Fig.\ref{fig:Schematics}A is an exemplary biological copying process where information of DNA sequence is copied by the DNA polymerase.   

When the average reaction currents along the kinetic path associated with correct and incorrect substrate incorporation to the copy strand are defined as $\langle J^c\rangle$ and $\langle J^i\rangle$, respectively, 
the error probability, which will be discussed throughout this paper, is given {\color{black}by the ratio of two reaction currents}  
\begin{align}
\eta=\frac{\langle J^i\rangle}{\langle J^c\rangle +\langle J^i\rangle}. 
\end{align} 
Error reducing strategies of biological copying processes are at work to minimize $\eta$ to a level acceptable for the survival of an organism. 

The mechanism of Bennett model (Fig.\ref{fig:Schematics}) \cite{Bennett1976} 
uses the chemical potential of substrates, whose concentrations are kept out of equilibrium ($|\Delta\mu|\gg|\Delta\mu_{eq}|$), as the free energy drive. 
In the model, correct and incorrect substrates are \emph{kinetically discriminated} with different kinetic barriers, but with no difference in binding stabilities of the two substrate types. 
At equilibrium, $\langle J^c\rangle=\langle J^i\rangle=0$, and the error rate ($f=\langle J^i\rangle/\langle J^c\rangle$) is solely determined by the ratio of equilibrium binding probabilities to copying system ($f_0=1$), so that $\eta_{eq}=f_0/(1+f_0)=1/2$. 
{\color{black}
When the free energy drive is large ($|\Delta\mu|  \rightarrow \infty $), 
the error rate converges to $\eta_0=1/(1+e^{\beta\delta})$, which is solely determined by the difference between the kinetic barriers for substrate binding, $\beta\delta$. Thus, as long as $\delta>0$, the mechanism can reduce the value of $\eta$ from $\eta_{eq}$ to $\eta_{0}$ at the expense of the free energy drive.}
See SI Text for the generalization of Bennett model where the equilibrium error rate is given by $f_0=e^{-\beta\Delta\mu_i}/e^{-\beta\Delta\mu_c}$.

Meanwhile, the original Hopfield model \cite{Hopfield1974} (see Fig.\ref{fig:Schematics}B) 
assumes that the binding rates (${\bf E}+{\bf C}\rightarrow {\bf EC}$ or ${\bf E}+{\bf I}\rightarrow {\bf EI}$ in Fig.\ref{fig:Schematics}B) for the correct and incorrect substrates are identical ($k^c_{\text{on}}[c]=k^i_{\text{on}}[i]$). 
In discriminating correct substrates from incorrect ones, the mechanism takes advantage of the facilitated unbinding of incorrect substrate from the copying system twice along the reaction path (${\bf EI}\rightarrow {\bf E}+{\bf I}$ and ${\bf EI}^*\rightarrow {\bf E}+{\bf I}$ in Fig.\ref{fig:Schematics}B, bottom), assisted by the extra free energy from molecular fuel consumption (GTP or ATP hydrolysis), which renders the reaction paths ${\bf EC}\rightarrow{\bf EC}^*$ and ${\bf EI}\rightarrow{\bf EI}^*$ effectively irreversible. 
The substrates complementary to the template polymer sequence are more likely to be polymerized, whereas the preferential unbinding of incorrect substrates from the copying complex end up with expending the energy for \emph{proofreading}, giving rise to the futile cycle.    
The mechanism of Hopfield model, called the \emph{kinetic proofreading mechanism}, reduces the error rate from $f_0$ down to $f_0^2$  \cite{Hopfield1974}. 

Real biological copying processes modify or combine the above two error-reducing strategies. 
More details on the different types of error reducing strategies and their combined effects can be found in refs.~\cite{Sartori2013,Pigolotti2016}.
\\

\begin{figure}[t]
%	\begin{center}
%\centering 
	\includegraphics[width=0.475\textwidth]{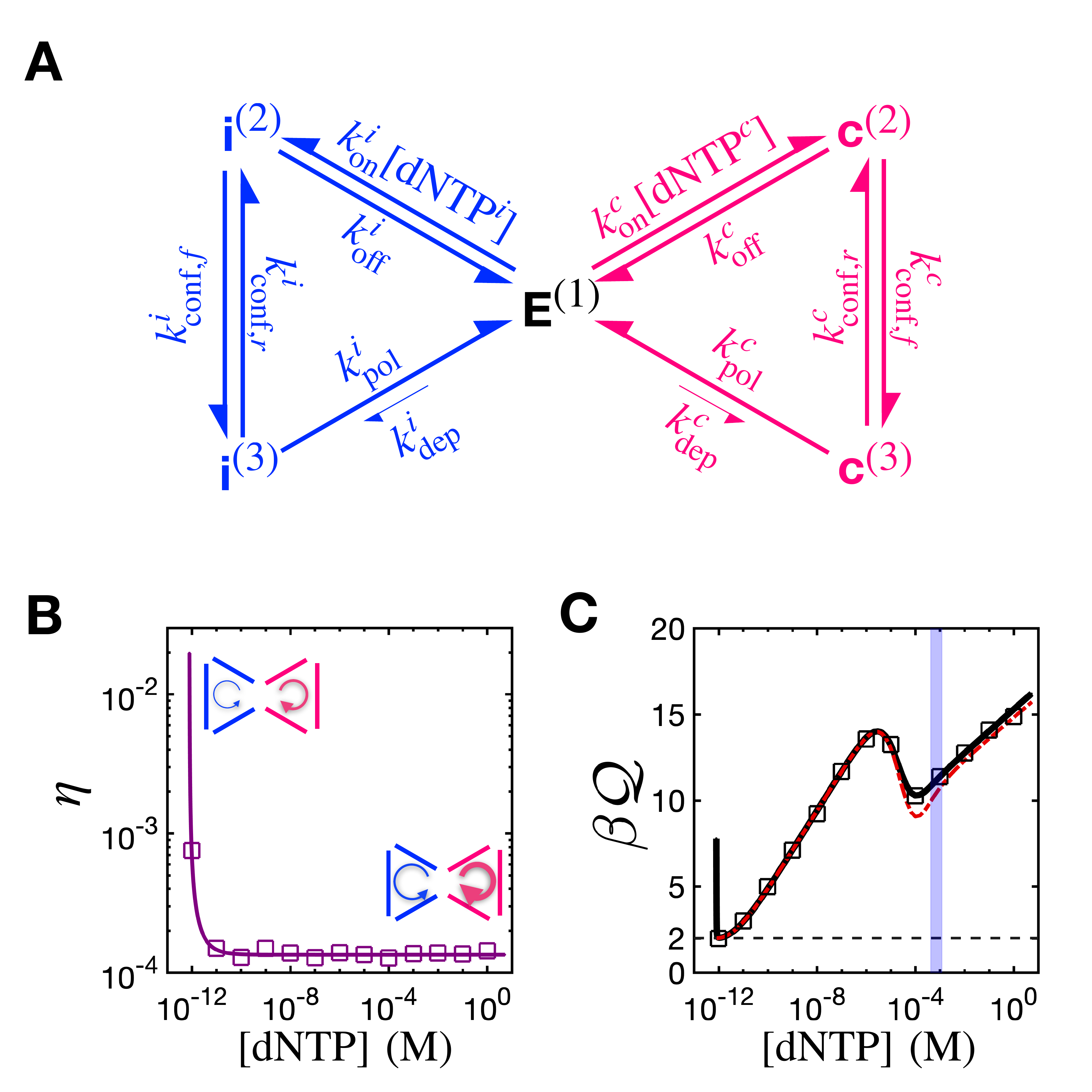}
	\caption{ The error reduction of DNA replication by the exonuclease-deficient T7 DNA polymerase.
	(A) The kinetic network for the dynamics of DNA polymerase \cite{Tsai2006}. 
	$[\text{dNTP}^c]$ and $[\text{dNTP}^i]$ are the concentration of the correct and incorrect nucleotides, respectively,
	where $[\text{dNTP}^i]=3[\text{dNTP}^c]$ holds from the assumption that all 4 substrates are present at identical concentrations.
	(B) The error probability ($\eta$)  as a function of $[\text{dNTP}] = [\text{dNTP}^c]+[\text{dNTP}^i]$. 
	With increasing [dNTP], relatively more reaction current flows in the subcycle associated with correct nucleotide incorporation. 
	(C) $\mathcal{Q}$ of T7 DNA polymerase as functions of  [dNTP]. 
	 The dash-dotted red line represents $\mathcal{Q}$ of an analogously defined unicyclic network with rate constants corresponding to the correct nucleotide incorporation pathway. 
	 The range of dNTP concentrations in \textit{E. coli} is demarcated with the purple shade \cite{Bochner1982,Buckstein2008,Schaaper2013}. 
	 For (B) and (C), the data points (squares) represent results from stochastic simulations using the Gillespie algorithm (see SI).
	 See Fig.~\ref{fig:PolSup} for other related dynamical properties. 	 
	 }
	\label{fig:PolPlot}
%	\end{center}
\end{figure}

{\bf Kinetic discrimination of dNTP by the T7 DNA polymerase. }
The DNA polymerase, in the absence of exonuclease activity, 
%, which carries out replication through its characteristic mode of conformational dynamics \cite{Raper2018}, 
is an enzyme that adapts the kinetic discrimination mechanism to reduce errors in replication    \cite{Tsai2006,Johnson2010,Cady2009}.  
%Since the original work by Johnson and coworkers, 
%common in the family of high fidelity DNA polymerases . 
In its simplest form, the replication dynamics of DNA polymerases can be represented by a double-cyclic reversible 3-state network consisting of two topologically identical subcycles for the incorporation of correct and incorrect nucleotides (Fig.\ref{fig:PolPlot}A).   
%In the following, we assumed that the polymerase is highly processive along DNA, which is justified by the tight binding between the polymerase and DNA \cite{Johnson1993}. 
Following the binding of the substrate (dNTP) ($[(1)\rightleftharpoons (2)]$), the polymerase on DNA undergoes conformational change ($[(2)\rightleftharpoons (3)]$). 
Finally, the effectively irreversible polymerization associated with dNTP incorporation ($[(3)\rightleftharpoons (1)]$)
with $k^c_{\text{pol}}\gg k^c_{\text{dep}}$ and $k^i_{\text{pol}}\gg k^i_{\text{dep}}$, completes the kinetic cycle.
The free energy difference between the binding of correct and incorrect nucleotides is approximately $\approx 5$ $k_BT$ \cite{Goodman1997}, which implies that the error probability at equilibrium is $\eta_{eq}\approx 7 \times 10^{-3}$.
In the presence of non-equilibrium drive, the conditions of $k^c_{\text{conf},f} \gg k^i_{\text{conf},f}$  and $k^c_{\text{pol}} \gg k^i_{\text{pol}}$, engendering much larger reaction current along 
${\bf E}^{(1)} \rightleftharpoons {\bf c}^{(2)} \rightleftharpoons {\bf c}^{(3)} \rightleftharpoons {\bf E}^{(1)}$ than that along ${\bf E}^{(1)} \rightleftharpoons {\bf i}^{(2)} \rightleftharpoons {\bf i}^{(3)} \rightleftharpoons {\bf E}^{(1)}$, allows DNA polymerases to reduce $\eta$ below $\eta_{eq}$ \cite{Tsai2006,Johnson2010}.
%employ the kinetic discrimination mechanism to reduce $\eta$ below $\eta_{eq}$ \cite{Tsai2006,Johnson2010}. 

As the total reaction current of polymerization, $\langle J_{\text{pol}} \rangle =\langle J^{ c}_{\text{pol}} \rangle +\langle J^{ i}_{\text{pol}}\rangle$, 
is a natural output observable accessible, for instance, in single molecule experiments \cite{abbondanzieri2005Nature,wen2008Nature,kaiser2011Science}, 
we calculate $\mathcal{Q}$ of DNA replication as (see Eq.\ref{eqn:TUR}, and Materials and Methods)
\begin{align}
\mathcal{Q} = \mathcal{A} \frac{ \langle \delta J_{\text{pol}}^2 \rangle}{ \langle J_{\text{pol}} \rangle }.
\end{align}
Alternatively, one could conceive choosing the current of correct sequence incorporation, $J^{ c}$, as the output variable; 
however, unlike that of $J_{\rm pol}$, 
the measurement of $J^{\rm c}$ requires the explicit knowledge of the DNA sequence being synthesized, 
which is not readily accessible to an experimental observer. 
{\color{black}
As long as $\eta$ is small, 
it is expected that $ \langle J^{ c} \rangle \approx \langle J_{\rm pol} \rangle$, and $\langle (\delta J^{ c})^2 \rangle \approx \langle (\delta J_{\rm pol})^2 \rangle$; thus, choosing $J^{ c}$ as the output variable instead of $J_{\rm pol}$ will not significantly alter the value of $\mathcal{Q}$. 
}

The free energy cost for a single step of polymerization (affinity, $\mathcal{A}$) can be written as 
\begin{align}
\beta\mathcal{A} &=
-\beta \left[ (1-\eta)\Delta\mu_{c} + \eta\Delta\mu_{i} \right]
-\eta \ln{\eta} - (1-\eta) \ln{(1-\eta)}\nonumber\\
&\equiv-\beta\Delta\mu+I,
\label{eqn:ribosomeA}
\end{align}
where $\Delta\mu_c$ and $\Delta\mu_i$ are the chemical potential difference along the 
correct and incorrect and polymerization cycles, respectively.
%Similar to the kinetic discrimination strategy of Bennet's model, 
%the free energy cost used to reduce $\eta$ stems from the chemical potentials of the substrates.
$\beta\mathcal{A}$ can be decomposed into the free energy gain ($-\beta\Delta\mu$) and the Shannon-entropy ($I$) arising from the chance of incorporating correct versus incorrect monomers in the copy strand. 
It is noteworthy that although $I\leq I_{\rm max} (=\ln{2})$ is usually small compared to $-\beta\Delta\mu$, it represents a fundamental thermodynamic property associated with stochastic copying processes (see Eq.~\ref{eqn:generalA}) \cite{Sartori2015}.

We explore how $\mathcal{Q}$ is affected when dNTP concentration ([dNTP]), which serves as a proxy for the chemical potential drive ($-\beta\Delta\mu$ in Eq.~\ref{eqn:A}), increases.  
We assume that the four types of dNTPs (A, G, C, T) are maintained in solution at equal concentrations, 
%For a given dNTP type, the concentration of incorrect nucleotides is always three times larger than that of correct nucleotide. 
and use experimentally determined kinetic rates of the exonuclease-deficient T7 DNA polymerase
to calculate $\eta$ and $\mathcal{Q}$ (see Table S1) \cite{Tsai2006}. 
With increasing [dNTP], 
the reaction current flows predominantly in one of the subcycles ($\langle J^{ c}\rangle\gg \langle J^{ i}\rangle$), and 
$\eta$ decreases monotonically to values consistent with experimental measurements~\cite{Cady2009,Kunkel1994}
(Fig.~\ref{fig:PolPlot}B);  
by contrast, $\mathcal{Q}$ displays non-monotonic variation (Fig.~\ref{fig:PolPlot}C). 
For $\mathcal{Q}$, two minima are identified, one at $\mathcal{Q}\approx 2$ $k_BT$, and the other at $\mathcal{Q}\approx 10$ $k_BT$ ($[\text{dNTP}]\approx 100$ $\mu$M), suggesting a complex interplay between the dissipation, current and its fluctuation. 
The suboptimal value of $\mathcal{Q}$ with respect to substrate concentration 
was also observed in models of transport motors \cite{Hwang2018JPCL}. 
Notably, the latter minimum is found near the range of the \emph{in vivo} [dNTP] in \textit{E. coli} ($430-1200$ $\mu$M \cite{Bochner1982,Buckstein2008,Schaaper2013}) (Fig.~\ref{fig:PolPlot}C). 

\begin{figure*}[t]
	\begin{center}
	\includegraphics[width=.95 \textwidth]{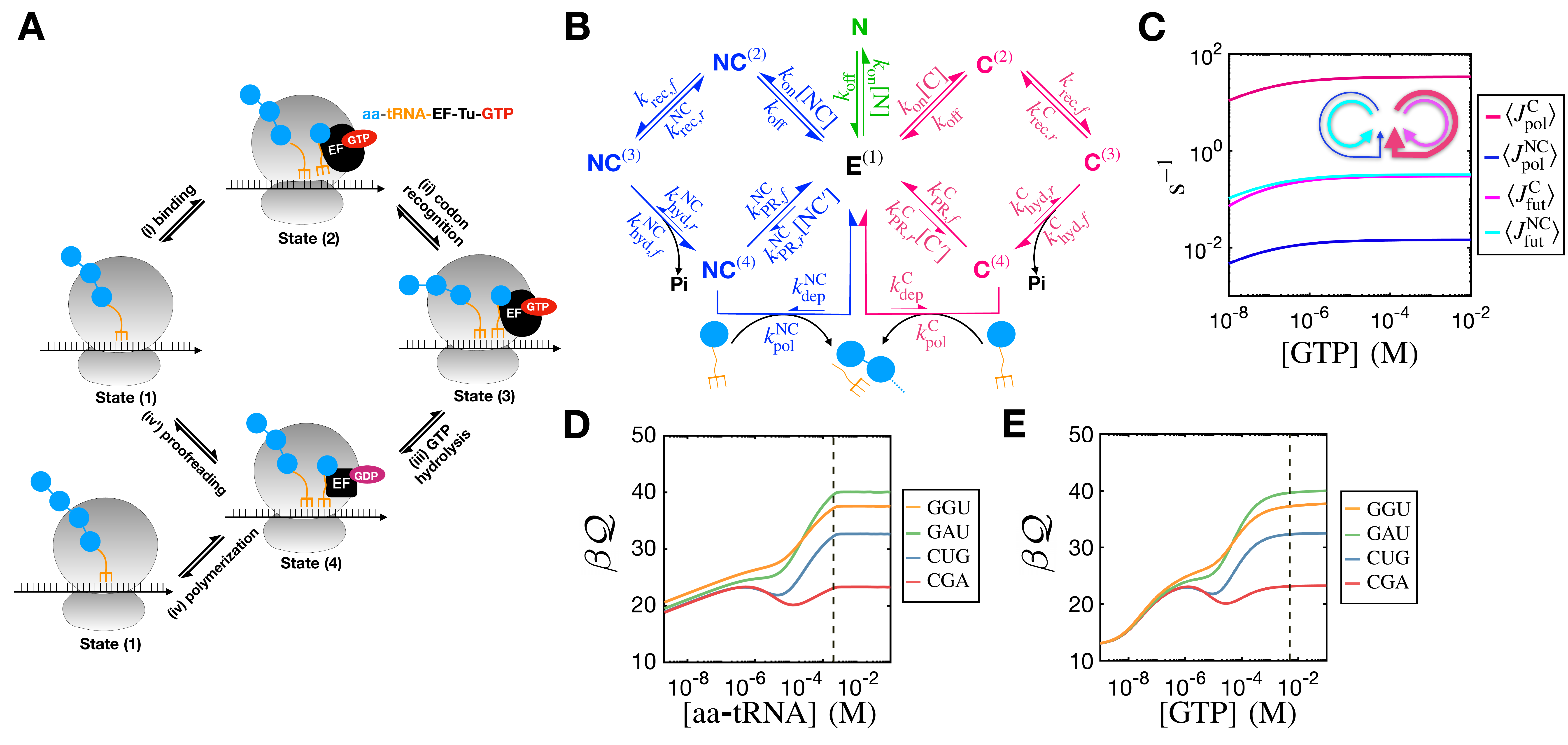}
	\caption{mRNA translation.
	(A)(B) Schematics of the catalytic cycle of the \textit{E. coli}  ribosome.
	In (B), ${\rm C}^{(i)}$ and $\text{NC}^{(i)}$ ($i=2,3,4$) represent intermediate states of the cognate and near-cognate aa-tRNA incorporation pathway.
	The state N represents the binding of the non-cognate aa-tRNA. 
	[C], [NC], and [N] represent the concentration of the cognate, near-cognate, and non-cognate ternary complex substrate, (aa-tRNA)-(EF-Tu)-GTP, respectively.
	$\rm [C']$ and $\rm [NC']$ represent the concentration of the cognate and near-cognate (aa-tRNA)-(EF-Tu)-GDP, respectively.
	(C) The currents along the kinetic cycles as a function of [GTP] for codon CUG.
	The thickness of the lines in the inset schematic represents the relative magnitude of the reaction currents: $\langle J^{{\rm C}}_{\text{pol}} \rangle \gg \langle J^{\text{NC}}_{\rm fut} \rangle \gtrsim \langle J^{{\rm C}}_{\rm fut} \rangle \gg \langle J^{\text{NC}}_{\text{pol}}\rangle $. 
	(D) $\mathcal{Q}$ as a function of [aa-tRNA].
	(E) $\mathcal{Q}$ as a function of [GTP].
	For (D) and (E), the dotted black line represents the cellular concentration in \textit{E. coli}.
	See Fig.~\ref{fig:RibosomeSup} for other related dynamical properties.
	}
	\label{fig:Ribosome}
	\end{center}
\end{figure*}

To understand the nature of the two minima of $\mathcal{Q}$, we calculated $\mathcal{Q}$ of an analogously defined unicyclic 3-state model with kinetic rates identical to those of the correct nucleotide incorporation cycle. 
The comparison between the $\mathcal{Q}$ of the two models suggests: 
(i) the global minimum is formed near the DB condition 
$\text{[dNTP]}\approx \frac{k^{c}_{\text{off}} k_{\text{conf},r}^c k^c_{\text{dep}}}{k^{c}_{\text{on}} k_{\text{conf},f}^c k^c_{\text{pol}}}$;  
(ii) the other minimum at $[\text{dNTP}]\approx 100$ $\mu$M arises from the Michaelis-Menten (MM) type enzyme kinetics. 
%Overall, it is noteworthy that the physiological dNTP concentrations that lead to the minimization of $\eta$ may also lead to the reduction of $\mathcal{Q}$. 
For Michaelis-Menten enzyme reactions, $\mathcal{Q}$ is suboptimized when the substrate concentration is near the Michaelis-Menten constant ($[S]\approx K_m$), where the response of the reaction is maximal with respect to the logarithmic variation of substrate concentration (see SI text). 
\\

{\bf Simplified model of mRNA translation. }
Since its introduction by Hopfield and Ninio \cite{Hopfield1974,Ninio1975}, 
kinetic proofreading has been the most extensively discussed error reducing strategy \cite{Banerjee2017,Murugan2012a,Rao2015,Pigolotti2016}. 
The proofreading reduces copy error by a resetting reaction that incurs an extra free energy.  
We study the effect of kinetic proofreading on $\mathcal{Q}$ by taking mRNA translation of the \textit{E. coli} ribosome as our model system (see Fig.~\ref{fig:Ribosome}).

The ribosome translates mRNA sequences into a polypeptide by reading \emph{codons}, each consisting of three consecutive nucleic acids (Fig.~\ref{fig:Ribosome}A). 
%For each codon, ribosome-mRNA complex accommodates aminoacyl-tRNAs (aa-tRNA) with the matching anti-codon sequence. 
When an aa-tRNA of a `matching' codon  binds to the ribosome-mRNA complex, 
the ribosome undergoes the reaction cycle for the \emph{cognate} aa-tRNA incorporation (red cycle in Fig.~\ref{fig:Ribosome}B). 
A \emph{near-cognate} aa-tRNA with a single mismatch can
also be incorporated, through a topologically identical but different kinetic pathway (blue cycle in Fig.\ref{fig:Ribosome}B).
For aa-tRNAs with two or three mismatches, corresponding to non-cognate aa-tRNAs, they can only interact with the ribosome-mRNA complex, but cannot undergo full incorporation (non-cognate aa-tRNA binding that corresponds to the reversible pathway colored in green in Fig.~\ref{fig:Ribosome}B) \cite{Dong1996,Wohlgemuth2011}. 
%For each codon, the concentration of cognate, near-cognate, and non-cognate aa-tRNAs are different \cite{Dong1996} (details given in the SI). 

Translation by the ribosome occurs via the following steps: 
(i) an aa-tRNA is accommodated to the ribosome-mRNA complex in the form of the (aa-tRNA)-(EF-Tu)-GTP complex [$(1)\rightleftharpoons (2)$], 
followed by (ii) the pairing of the codon-anticodon sequence $[(2)\rightleftharpoons (3)]$. 
(iii) GTP hydrolysis and the conformational change of EF-Tu $[(3)\rightleftharpoons (4)]$. 
(iv) A new peptide bond formation with the ribosome translocating to the next codon ($k_{\rm{pol}}^{{\rm C}}$ and $k_{\rm{pol}}^{\text{NC}}$), 
or 
(iv$^\prime$) dissociation of (aa-tRNA)-(EF-Tu)-GDP complex from the ribosome (i.e. $k_{{\rm{PR}},f}^{{\rm C}}$ and $k_{{\rm{PR}},f}^{\text{NC}}$).
Both steps of (iv) and (iv$^\prime$) reset the system back to the state (1) $[(4)\rightleftharpoons (1)]$.
The cognate aa-tRNAs are differentiated from near-cognate aa-tRNAs mainly due to the faster rates of GTP hydrolysis and peptide bond formation ($k^{{\rm C}}_{\text{hyd},f} \gg k^{\text{NC}}_{\text{hyd},f}$ and $k^{{\rm C}}_{\text{pol}}\gg k^{\text{NC}}_{\text{pol}}$). 
The rates associated with tRNA binding, unbinding and recognition are similar between the two. 
%The binding and unbinding rates of all the aa-tRNAs are equal, and the forward rates of codon-recognition for the cognate and near-cognate aa-tRNAs are also equal ($k^{{\bf c}}_{\text{rec},f}=k^{\text{NC}}_{\text{rec},f}$).
%However, the GTP hydrolysis and the peptide bond formation rates of the cognate aa-tRNA are faster than those of the near-cognate aa-tRNA, respectively . 
As a result, the reaction current of incorporating the cognate aa-tRNA is greater than that of the near-cognate aa-tRNA along the network depicted in Fig.~\ref{fig:Ribosome}B. 
Because the incorporation current of non-cognate aa-tRNA is effectively zero ($\langle J_{\text{pol}}^{\rm N} \rangle=0$), 
the error probability of the ribosome is  $\eta = \langle J^{\text{NC}}_{\text{pol}} \rangle/(\langle J^{{\rm C}}_{\text{pol}} \rangle+\langle J^{\text{NC}}_{\text{pol}}\rangle)$, where $\langle J^{{\rm C}}_{\text{pol}} \rangle$ and $\langle J^{\text{NC}}_{\text{pol}} \rangle$ are the currents of cognate and near-cognate aa-tRNA incorporations, respectively.

%Selecting the total reaction current of polymerization, $\langle J_{\text{pol}} \rangle =\langle J^{\rm C}_{\text{pol}} \rangle +\langle J^{\rm NC}_{\text{pol}}%\rangle$, as the output observable, we calculate $\mathcal{Q}$ of mRNA translation as 
%$\mathcal{Q} = \mathcal{A} \frac{ \langle \delta J_{\text{pol}}^2 \rangle}{ \langle J_{\text{pol}} \rangle }$,
Similar to DNA replication, the free energy cost for a single step of translation ($\mathcal{A}$) can be written as
\begin{align}
\beta\mathcal{A} =&
-\beta \left[ \Delta\mu_{\text{pol}} + \frac{ \langle J_{\rm fut} \rangle}{\langle J_{\text{pol}} \rangle}\Delta\mu_{\rm fut} \right]
-\eta \ln{\eta} - (1-\eta) \ln{(1-\eta)}. 
\label{eqn:ribosomeA}
\end{align}
Here, $\Delta\mu_{\rm fut}$ and $\Delta\mu_{\text{pol}}$ are the chemical potential difference along the 
futile and polymerization cycles, respectively (see SI for details). 
The kinetic proofreading uses extra energy in the form of GTP hydrolysis ($\Delta\mu_{\rm fut}$), engendering futile cycles, and reduces $\eta$ 
further than that by kinetic discrimination alone, the latter of which only capitalizes on the thermodynamic cost of polymerization ($\Delta\mu_{\text{pol}}$).

The dynamics of mRNA translation was examined as a function of the concentration of aa-tRNA and GTP by assuming that the ternary complex concentration was in pseudo-equilibrium 
with respect to the concentration of its components, aa-tRNA, EF-Tu, GTP, and GDP (see SI). 
%Briefly, the ternary complex substrates for the initial binding reactions ([C], [NC], and [N] in Fig.~\ref{fig:Ribosome}B) increase with GTP, 
%while those of the reverse proofreading reactions ($\rm[C']$ and $\rm[NC']$ in Fig.~\ref{fig:Ribosome}B) generally decrease as a function of GTP (Fig.~\ref{fig:GTPdependence}).
With increasing [GTP], the polymerization current of all cycles increases
while maintaining their relative magnitudes: $\langle J^{{\rm C}}_{\text{pol}} \rangle \gg \langle J^{\text{NC}}_{\rm fut} \rangle \gtrsim \langle J^{{\rm C}}_{\rm fut} \rangle \gg \langle J^{\text{NC}}_{\text{pol}}\rangle $ (Fig.~\ref{fig:Ribosome}C). 
In other words, while most cognate aa-tRNAs that reach state ${\bf C}^{(4)}$ are polymerized, 
most of the near-cognate aa-tRNAs that reach state ${\bf NC}^{(4)}$ are rejected by the proofreading reaction.

%Even at vanishing GTP concentrations, the system does not reach the DB condition,
%since the WT level of GDP (.6 mM) is high enough to drive currents along $\rm E^{(1)}\rightleftharpoons C^{(2)} \rightleftharpoons C^{(3)}\rightleftharpoons C^{(4)} \rightleftharpoons E^{(1)}$ and $\rm E^{(1)}\rightleftharpoons NC^{(2)} \rightleftharpoons NC^{(3)}\rightleftharpoons NC^{(4)} \rightleftharpoons E^{(1)}$.

For all codon types, $\eta$ is nearly constant for a wide range of [aa-tRNA] and [GTP] (Figs.~\ref{fig:RibosomeSup}A, E).
In contrast, the shape of $\mathcal{Q}$ varies depending on the codon (Figs.~\ref{fig:Ribosome}D, E). 
For most codons, $\mathcal{Q}$ increases monotonically with [aa-tRNA] and [GTP]. 
For codons CGA and CUG, $\mathcal{Q}$ has a local minimum at [aa-tRNA]$\approx10$ $\mu$M and [GTP]$\approx$ $10$ $\mu$M. 
The distinguishing feature of the codons CGA and CUG is their high cognate to near-cognate aa-tRNA concentration ratios (${\rm [C]/[NC]} \approx 0.9$ for CGA and ${\rm [C]/[NC]}\approx0.5$ for CUG. Fig.~\ref{fig:codon_table}),
which suggests that the local minimum of $\mathcal{Q}$ occurs when the contribution from the near-cognate incorporation pathway is relatively low.
As seen in the case of T7 DNA polymerase (Fig.~\ref{fig:PolPlot}C and Fig.~\ref{fig:PolSup}), the local minimum of $\mathcal{Q}$ (Figs.\ref{fig:Ribosome}D, E), if any, is identified at regions where the response of $\langle J_{\text{pol}}\rangle$ is large with respect to the logarithmic variation of [aa-tRNA] or [GTP] (Figs.~\ref{fig:RibosomeSup}B, F). 
\\

\begin{figure*}[t]
	\begin{center}
	\includegraphics[width=.95 \textwidth]{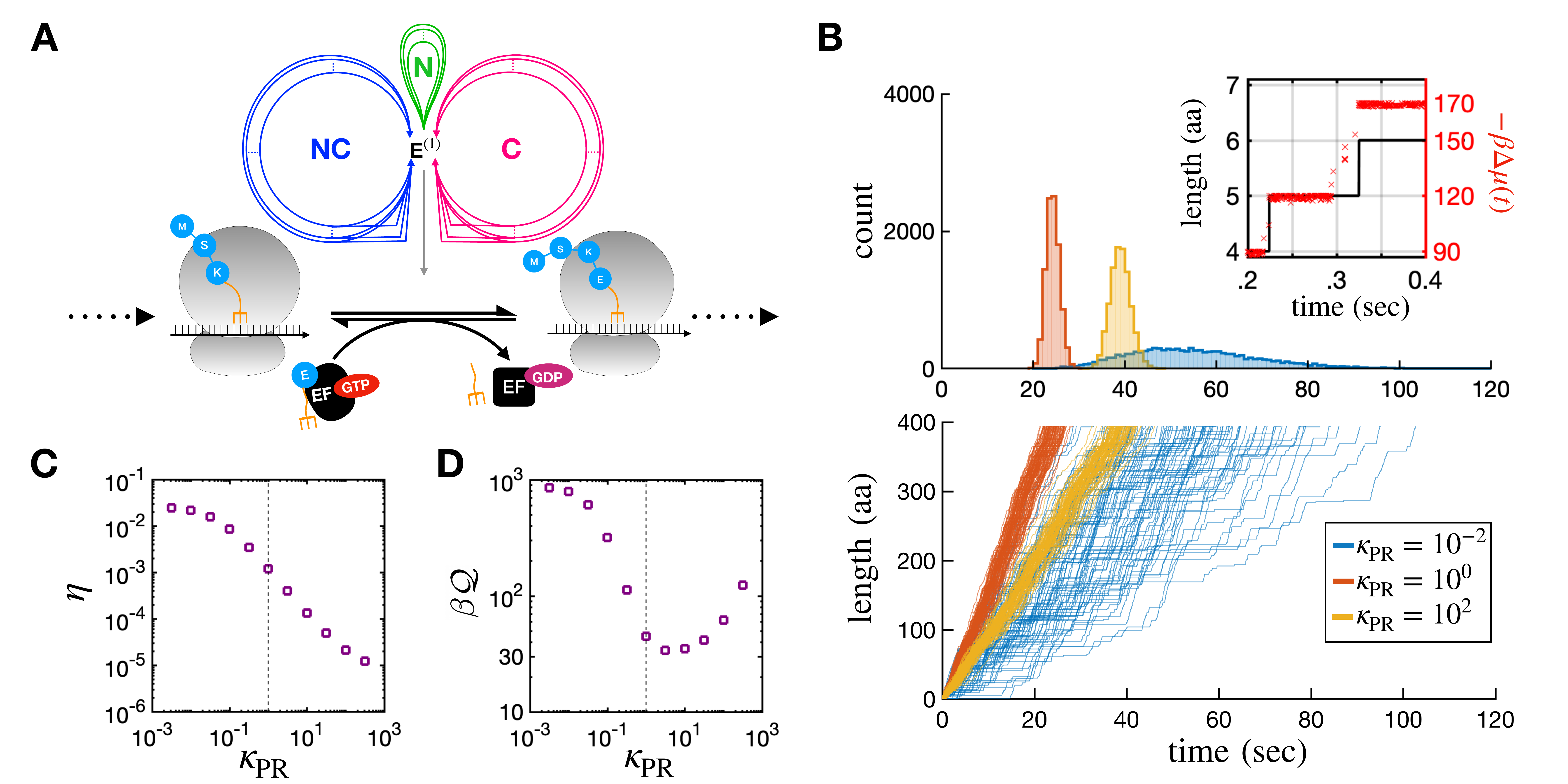}
	\caption{
	The reaction kinetics of translation with 42 aa-tRNA species. 
	(A) Schematic of the translation of the \textit{tufB} mRNA sequence into EF-Tu. For each reaction cycle, 42 different aa-tRNAs can bind to the apo state of the enzyme. Out of these, the cognate and near cognate aa-tRNAs can undergo the reaction cycle previously defined in Figure~\ref{fig:Ribosome}B. For more detail on the simulated reaction network, refer to the SI. 
	(B) (Bottom) An ensemble of time traces ($N=100$) generated from the numerics using Gillespie algorithm that simulates the mRNA translation (or the synthesis of the polypeptides consisting of 394 amino-acids) at different values of $\kappa_{\text{PR}}$.
              (Top) The histogram of translation completion times. 
            The inset shows a sample trajectory at the wild type condition ($\kappa_{\text{PR}}=1$), at which a proofreading reaction occurs at around 0.3 seconds. 
            The average dissipation from the process is shown with red crosses ($\Delta\mu(t)$). 
            Whenever the proofreading takes place, the synthesis of polypeptide is stalled. 
        (C) The error probability and (D) $\mathcal{Q}$ of TUR are plotted against $\kappa_{\text{PR}}$. 
        The dotted lines depict the wild type condition ($\kappa_{\text{PR}}=1$). 
	 }
	\label{fig:Translation}
	\end{center}
\end{figure*}

{\bf Multicyclic model of mRNA translation. }
To address the mRNA translation in a more realistic fashion, 
we consider a multicyclic model which translates 42 species of aa-tRNAs into 20 different amino-acids (Fig.~\ref{fig:Translation}).   
For each codon, the 42 aa-tRNAs are grouped into cognate, near-cognate and non-cognate types (Fig.~\ref{fig:codon_table}). 
Using the information on the concentration of 42 aa-tRNAs and the  model illustrated in Fig.~\ref{fig:Translation}A, we simulated the translation of the \textit{tufB} mRNA sequence consisting of $n_{\text{aa}}=394$ amino-acids, which encodes for EF-Tu, 
a highly abundant protein in \textit{E. coli} \cite{Ishihama2008} (Fig.~\ref{fig:Translation}).

The dynamics arising from the multicyclic model are studied using an ensemble of trajectories generated from Gillespie simulations (Fig.~\ref{fig:Translation}B). % generated with different values of $\kappa_{\text{PR}}$ (Fig.~\ref{fig:Translation}B). 
The total number of translational steps ($N_{tln}$) 
that complete the polymerization of the full amino-acid sequences varies from one realization to another.
%as each realization of the process would accommodate cognate or near-cognate amino-acids via entirely different reaction paths.   
%Next, let $x^{s}_{1},\dots,x^{s}_{N_s}$ be the sequence of Markovian steps occurring at the respective sequence of time points $t^s_{1},\dots,t^{s}_{N_s}$. 
Selecting the completion time of translation ($\mathcal{T}$) as the output observable for each dynamic process, we  define TUR of translation as
\begin{align}
\mathcal{Q}
= \left[ -\Delta\mu + \beta^{-1}I \right] \frac{ \langle (\delta \mathcal{T})^2 \rangle}{ \langle \mathcal{T} \rangle^2 },
\end{align}
where, similar to all previous models, the dissipation has contributions from the free energy drive ($ \Delta\mu$) and Shannon-entropy ($I$). 
Denoting the forward and  reverse rate constants of each kinetic step by $k_{i,f}$ and $k_{i,r}$ for $i=1,\dots,N_{tln}$, we can compute the average free energy drive by
$-\beta \Delta\mu  =\Big\langle\sum_{i=1}^{N_{tln}} \ln{(k_{i,f}/k_{i,r})}\Big\rangle$, where $\langle\ldots\rangle$ denotes the average over the ensemble of $10^4$ realizations. 
The entropic contribution can be computed as 
$I = -\sum_{l=1}^{n_{\text{aa}}}\sum_{i_{\text{aa}}=1}^{20} \eta^l_{i_\text{aa}} \ln{\eta^l_{i_\text{aa}}},
%\label{eqn:extendedI}
$
where $\eta^l_{i_\text{aa}}$ is the probability of incorporating one of the 20 types of amino-acids, at the $l$-th position. 
%It is noteworthy that although $I\leq I_{\rm max} (=n_{\text{aa}}\ln20)$ is small compared to $-\beta\Delta\mu$, it represents a fundamental thermodynamic property associated with stochastic copying processes. 
% and is essential for the correct expression of $ q $. 
%For instance, without the $I$ term, a stochastic simulation of the Bennett model with $k_f^c=k^r_c$ and $k^i_f=k^i_r$ would yield net forward motion even in the absence of dissipation!

Using the multicyclic model, we evaluated $\eta$ and $\mathcal{Q}$ with respect to perturbations to the proofreading reaction, 
by considering a multiplication factor $\kappa_{\text{PR}}$ to the original wild-type (WT) rate constants $k^{{\rm C}}_{\text{PR},f}$, $k^{{\rm C}}_{\text{PR},r}$, $k^{\text{NC}}_{\text{PR},f}$, and $k^{\text{NC}}_{\text{PR},r}$.
Although the rate constants are not experimentally tunable parameters like $\rm [GTP]$, the cell can optimize them throughout the  evolution by means of mutations to the ribosome, EF-Tu, and tRNA.  
This type of perturbative analysis can be used to decipher which feature of the reaction kinetics for mRNA translation is optimized in the cell (see the effect of other perturbations in Fig.~\ref{fig:RibosomeSup3}).

The WT level of proofreading gives rise to an average speed $\langle J_{\text{pol}} \rangle \approx$16 aa/sec and error probability $\eta \approx 10^{-3}$ in our simulation, consistent with the experimental measurements \cite{Bouadloun1983, Young1976}. 
While $\eta$ decreases monotonically with $\kappa_{\text{PR}}$, $\mathcal{Q}$ is non-monotonic with $\kappa_{\text{PR}}$, minimized near the wild type condition. 
At $\kappa_{\text{PR}} = 1$ we obtain $\mathcal{Q} \approx 45$ $k_BT$ \cite{Pineros2020} (Fig.~\ref{fig:Translation}D). For the given kinetic parameters from WT, $\mathcal{Q}$ is minimized to $\sim$ 30 $k_BT$ when the rates of proofreading is augmented by 5 fold. 
In a scenario of negligibly low proofreading ($\kappa_{\text{PR}} = 10^{-2}$), 
the completion times for the translation display a much broader distribution than that by the WT 
($\kappa_{\text{PR}} = 1$). 
Thus, near the WT condition, proofreading can simultaneously improve the fidelity of translation and suppress the fluctuation of protein synthesis in an energetically efficient way.

Importantly, fluctuations in the completion time for mRNA translation can be critical, 
as it could in turn lead to significant variation in protein copy number. 
Thus, our results demonstrate that kinetic proofreading, an error reducing strategy, can also contribute to the energetically efficient control of protein levels.

\section{Discussion}
{\bf Implications of the T7 DNA polymerase model. }
In the wild type T7 DNA polymerase,
the proofreading activity of the exonuclease further reduces $\eta$ by two orders of magnitude \cite{Donlin1991}.
In fact, 
in more complex systems such as DNA replication of \textit{E. coli},
the combination of the actions of DNA polymerase, 
exonuclease, and mismatch repair machineries 
achieves an error probability as small as $\eta\approx 10^{-10}$ \cite{Schaaper1993}. 
Although these extra components of DNA replication
could in principle be included in our model \cite{Bennett1976,banerjee2017PNAS,Gaspard2016a,Hoekstra2017a}, 
general consensus on their kinetic network and measurement of kinetic rates are currently lacking.  
Thus, we focused on the simpler, yet still experimentally realizable, exonuclease-deficient T7 DNA polymerase,
which has served as a useful tool for sequencing technologies and for biochemical studies of DNA polymerases  \cite{Zhu2014,Tsai2006}.

For the exonuclease-deficient T7 DNA polymerase, 
we found that $\mathcal{Q}$ is suboptimized near the physiological [dNTP]. 
Similarly, it has recently been discovered that in metabolic reactions, 
the physiological substrate concentrations are generally close to their respective $K_m$ values \cite{Park2016}.
A systems level analysis of yeast metabolism also showed that 
reaction currents of metabolism are generally self-regulated to the values at which their response to the change in substrate concentration is significant \cite{Hackett2016}.
In light of our analysis of Michaelis-Menten enzyme reactions (see the section {\bf The suboptimal condition of reversible Michaelis-Menten reactions} in SI), 
the above-mentioned condition of metabolism is closely related with the condition of suboptimized $\mathcal{Q}$. 
\\

\begin{figure}[t]
	\includegraphics[width=0.475 \textwidth]{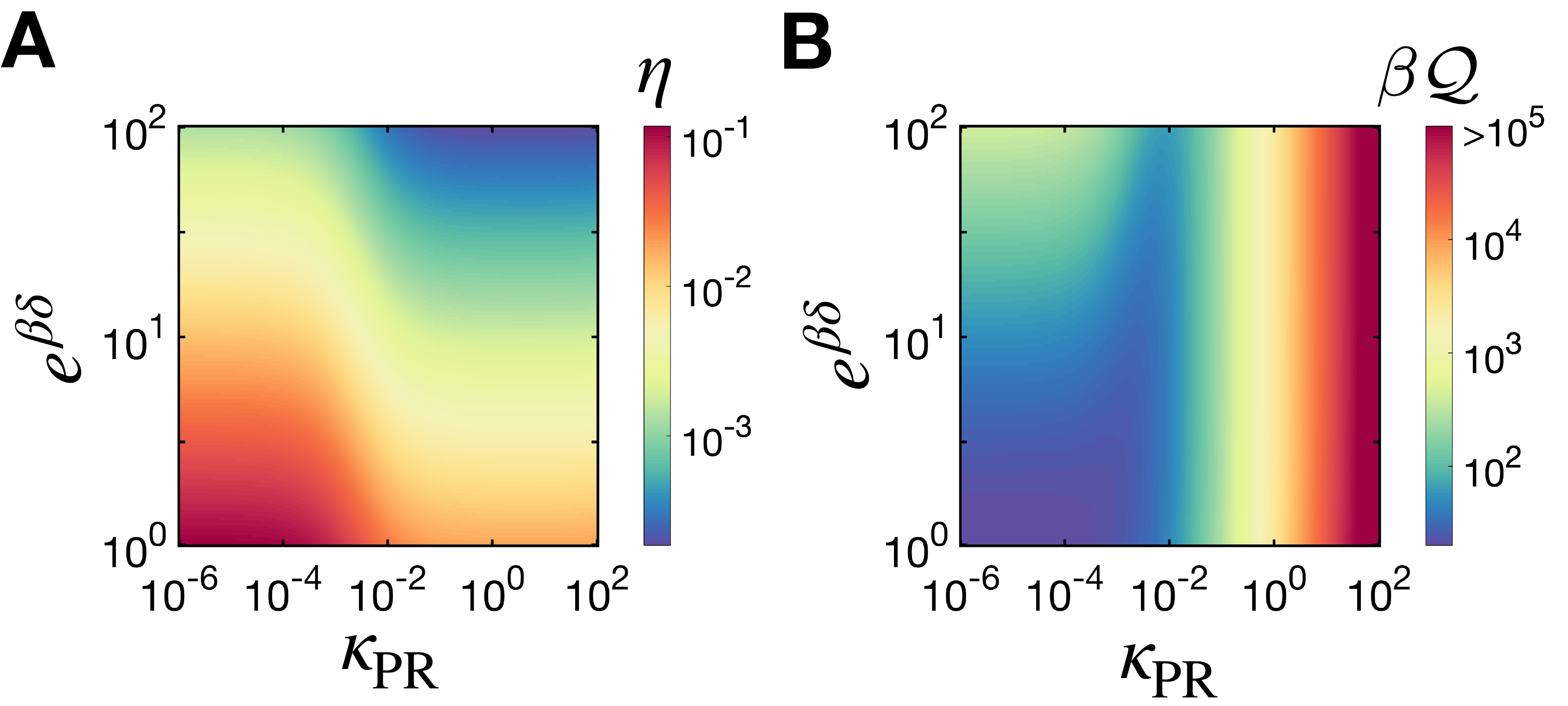}
	\caption{
	The modified Hopfield model with kinetic discrimination.
%	(A) Schematic of the modified Hopfield model with kinetic discrimination.
	(A) The error probability ($\eta$) and (B) $\mathcal{Q}$ with respect to variations in $\delta$ and $\kappa_{\text{PR}}$ defined in the main text. 
	The rate constants used to generate the plots are given in Table S3. 
	See Fig.~\ref{fig:HopfieldSI} for other related dynamical properties plotted for $\beta\delta=0$ and $\beta\delta=\ln 10$.
	}
	\label{fig:Hopfield}
\end{figure}

{\bf mRNA translation combines the strategies of kinetic discrimination and proofreading. }
The non-monotonic variation of $\mathcal{Q}$ with $\kappa_{\text{PR}}$ (Fig.~\ref{fig:Translation}D) is not a feature of the original kinetic proofreading model, which lacks the forward kinetic discrimination (i.e. $\beta\delta=0$). 
As the perturbative parameter $\kappa_{\text{PR}}$ is increased, the error rate ($f=\langle J^i\rangle/\langle J^c\rangle$) is reduced to $f \gtrsim f_0^2 = e^{-2\beta\left(\Delta \mu_i - \Delta \mu_c\right)}$ (Fig.~\ref{fig:HopfieldSI}A, blue line). 
Furthermore, in the original Hopfield model, $\lambda\approx 1$ regardless of $\kappa_{\text{PR}}$ (Fig.~\ref{fig:HopfieldSI}D), which leads to $\mathcal{Q} \approx \mathcal{A}$ (Fig.~\ref{fig:HopfieldSI}C, E), and 
a monotonically increasing $\mathcal{Q}$ with $\kappa_{\text{PR}}$ (Fig.~\ref{fig:Hopfield}C, and Fig.~\ref{fig:HopfieldSI}E).

To introduce the kinetic discrimination to the Hopfield model,
we consider a modified version, the associated kinetic constants of which satisfy the following relations with $\beta\delta>0$: 
\begin{align}
e^{\beta\delta}=\frac{k^c_{\text{on} }}{k^i_{\text{on}}} = \frac{k^c_{ {\text{hyd}},f }}{k^i_{{\text{hyd}},f}} =\frac{k^c_{\text{pol} }}{k^i_{\text{pol}}}= \frac{k^c_{ {\text{PR}},r }}{k^i_{ {\text{PR}},r }}. 
\end{align} 
As expected, $\eta$ decreases monotonically with $\beta\delta$ and $\kappa_{\text{PR}}$ (Fig.~\ref{fig:Hopfield}B). 
Qualitatively similar to mRNA translation, $\mathcal{Q}$ is minimized over a certain range of $\kappa_{\text{PR}}$ as long as $e^{\beta\delta}\gtrsim 10^1$ (Fig.~\ref{fig:Hopfield}B and Fig.~\ref{fig:HopfieldSI}E).
Taken together with the modified Hopfield model, 
mRNA translation in \textit{E. coli} balances the kinetic discrimination and proofreading, to attain low $\eta$ and suboptimized $\mathcal{Q}$.
\\

{\bf Optimality of the speed and TUR in the \textit{E. coli} ribosome. }
Similarly to our analysis shown in Fig.~\ref{fig:Translation}D, recent theoretical studies on mRNA translation by the ribosome \cite{Banerjee2017,Mallory2019} 
have also observed that while the error probability is still far from its minimum, the WT value of the mean first translation time ($\langle \tau_{\text{MFPT}}\rangle$) is close to its minimum; and hence it was concluded that the \textit{E. coli} ribosome is primarily optimized for speed. 
As far as the $\kappa_{\text{PR}}$-dependencies of speed ($\langle J_{\text{pol}} \rangle \approx \langle \tau_{\text{MFPT}}\rangle^{-1}$ ) and $\eta$ are concerned, 
our study points to the same finding (Fig.~\ref{fig:RibosomeSup3}). 
In fact, recent studies, which showed translational pausing caused protein misfolding, lend support to the significance of optimal codon translation speed \cite{nedialkova2015Cell,trovato2017BJ}.

Fast codon translation speed, small fluctuations in total translation time, and low thermodynamic costs could be favorable characteristics of translation, all likely under evolutionary selection pressure~\cite{Ilker2019PRL}; however, not all of these requirements can be fulfilled simultaneously.  
In this aspect, of great significance is our finding that the TUR measure of \textit{E. coli} ribosome ($\mathcal{Q}\approx 45$ $k_BT$) for the wild type condition is in the vicinity of its minimum with respect to $\kappa_{\text{PR}}$ ($\sim 30$ $k_BT$)  (Fig.~\ref{fig:Translation}D).  
\\

{\bf Significance of small $\mathcal{Q}$. }
The theoretical lower bound of TUR ($\mathcal{Q}=2k_BT$) allows us to endow physical significance to the  $\mathcal{Q}$ values obtained for the two essential copy machines ($\mathcal{Q}\approx10$ $k_BT$ for the T7 DNA polymerase and $\mathcal{Q}\approx45$ $k_BT$ for the \textit{E. coli} ribosome). 
For instance, we can compare $\mathcal{Q}$ of copying enzymes to molecular clocks, in which TUR is defined with respect to the tradeoff between the energetic cost and the uncertainty in the cycle duration. 
Marsland \emph{et al.} have recently demonstrated that TUR of multiple types of biochemical oscillators severely underperform the $2k_BT$ bound \cite{Marsland2019}. 
For the circadian KaiABC oscillator system, 
$\beta\mathcal{Q}\gtrsim \mathcal{O}(10^2)$. 
This either implies that the precision of cycle periodicity is the key priority over the energy expenditure, or that this synthetic biochemical cycle is not optimally designed under the constraint of TUR.    
In contrast, biological motors that transport cargo along cytoskeletal filaments display small $\beta\mathcal{Q} (\approx 7-15)$, simultaneously minimizing energetic costs, fluctuation, and maximizing speed \cite{Hwang2018JPCL}.  
Compared to biological motors harnessing the thermal fluctuations along with the ATP hydrolysis free energy,  
synthetic nanomachines \cite{kudernac2011Nature}, which uses $\sim eV$ UV-light source as the driving force, are expected to have much greater $\mathcal{Q}$ values.
While the biological function of copying enzymes is to maintain low copying error, 
it is remarkable to discover that T7 DNA polymerase and \textit{E. coli} ribosome are also working at conditions close to the theoretical bound dictated by the TUR.

\section{Methods}
When the number of steps taken by the enzyme is selected as the output observable ($X(t)=n(t)$ in Eq.\ref{eqn:TUR}), 
TUR in Eq.\ref{eqn:TUR} is modified to 
\begin{align}
\mathcal{Q}= q(t)\frac{\langle\delta n(t)^2\rangle}{\langle n(t)\rangle^2}= \mathcal{A} \lambda \geq 2 k_BT,
\end{align}
where $\mathcal{A}=q(t)/\langle n(t)\rangle$ and $\lambda = \langle \delta n(t)^2 \rangle/\langle n(t)\rangle $ is the Fano factor of the copying process, which can also be written as $\lambda=\langle \delta J^2\rangle/\langle J\rangle$.

\begin{acknowledgements}
This work was supported by the KIAS Individual Grant No. CG067102 (Y.S.) and No. CG035003 (C.H.) at Korea Institute for Advanced Study.  
We thank the Center for Advanced Computation in KIAS for providing computing resources.
\end{acknowledgements}

% Bibliography
\bibliography{library,mybib1}

% Figures

\title{Supporting Information: Thermodynamic cost, speed, fluctuations, and error reduction of biological copy machines}
% Please give the surname of the lead author for the running footer

% Please include corresponding author, author contribution and author declaration information
%\authorcontributions{Author contributions: Y.H.S. and C.H. designed research; Y.H.S. and C.H. performed research; Y.H.S. and C.H. analyzed data; and Y.H.S. and C.H. wrote the paper.
%}
%\authordeclaration{The authors declare no competing interest.}
%\equalauthors{\textsuperscript{1}A.O.(Author One) and A.T. (Author Two) contributed equally to this work (remove if not applicable).}
%\correspondingauthor{\textsuperscript{1}To whom correspondence should be addressed. E-mail: hyeoncb@kias.re.kr}

% Keywords are not mandatory, but authors are strongly encouraged to provide them. If provided, please include two to five keywords, separated by the pipe symbol, e.g:

\clearpage 

\setcounter{equation}{0}
\setcounter{figure}{0}
\renewcommand{\thetable}{S\arabic{table}}
\renewcommand{\thefigure}{S\arabic{figure}} 
\renewcommand{\theequation}{S\arabic{equation}}

\section{Supporting Information}
\subsection*{Bennett model: kinetic discrimination without proofreading}
\begin{figure}[t]
	\begin{center}
	\includegraphics[width=0.45\textwidth]{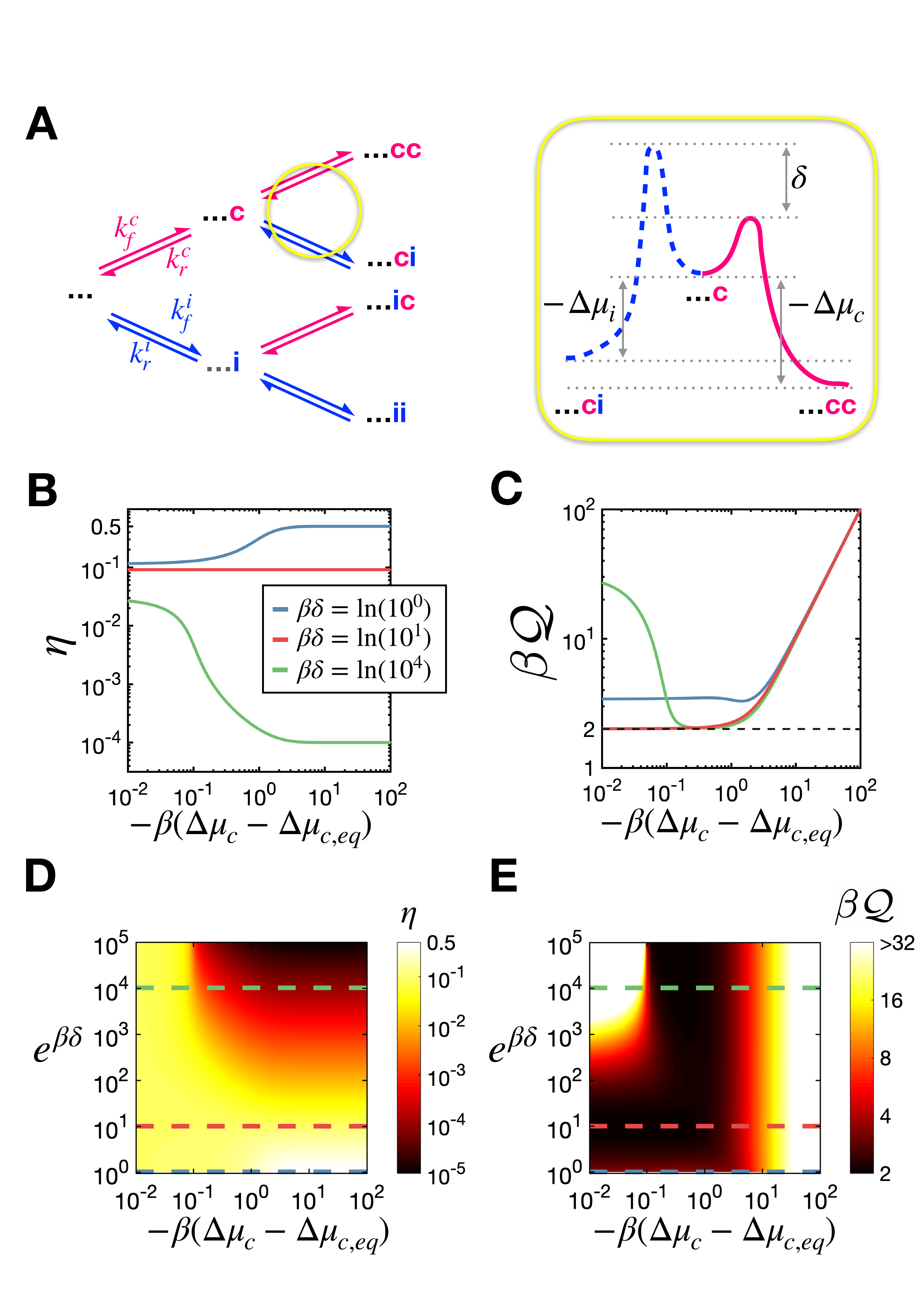}
	\caption{The reaction kinetics of a non-proofreading copying system generalizing Bennett model. 
	{(A)} (Left) The schematic of the copy polymer elongation. 
			The ($\dots$) represents the previously synthesized copy polymer. 
	(Right) The schematic of the free energy profile of monomer incorporation. 
			The free energies of incorporating correct and incorrect monomers are $\beta\Delta\mu_c =- \ln{(k^c_f/k^c_r)}$ and $\beta\Delta\mu_i=-\ln{(k^i_f/k^i_r)}$, respectively.
			The incorporation of the correct monomer occurs faster than that of the incorrect monomer, as described by the parameter $\beta\delta = \ln{(k^c_f/k^i_f)}>0$. (B) The error probability ($\eta$) and (C) $\mathcal{Q}$ as functions of the chemical potential bias for different values of the kinetic discrimination parameter ($\beta\delta$), where $e^{-\beta(\Delta\mu_c-\Delta\mu_i)}=10$.  
			The curves in {(C)} for various $\beta\delta$ values are color-coded identically as those in {(B)}. 
			In {(C)}, the physical limit of $\beta\mathcal{Q}=2$ is marked with a dashed line. 
			Diagrams of (D) $\eta$ and (E) $\beta\mathcal{Q}$ as a function of $e^{\beta\delta}$ and $-\beta(\Delta\mu_c-\Delta\mu_{c,eq})$. 		
	}
	\label{fig:BennettSchematicSI}
	\end{center}
\end{figure}

In the non-proofreading model of copy processes introduced by Bennett \cite{Bennett1976}, 
a copying enzyme synthesizes the complementary polymer strand by incorporating the monomers from the solution via a single kinetic step (Fig.~\ref{fig:BennettSchematicSI}A). 
The copy process is described by three key parameters: the binding free energies of correct and incorrect monomers, $\Delta\mu_c$ and $\Delta\mu_i$, and the difference in the kinetic barriers, $\delta$. 
Here, we keep the difference between the binding free energies ($\Delta\Delta \mu_{ci} \equiv \Delta\mu_c-\Delta\mu_i$) and the difference between the kinetic barriers constant. 
For instance, in DNA replication, $\Delta\Delta \mu_{ci}$ and $\delta$ 
are determined by the molecular properties of the nucleotides and the polymerase.
The error probability can be modulated by changing nucleotide concentrations, which corresponds to changing $\Delta\mu_c$ in the Bennett model. Thus, in the Bennett model, $\eta$ is evaluated as a function of $\Delta\mu_c$, and vice versa.
In the following, we derive the expressions of $\Delta\mu_c(\eta)$, $\lambda(\eta)$, $\mathcal{A}(\eta)$ and $\mathcal{Q}(\eta)$. 
Using these expressions, we plot the diagrams of $\eta$ and $\mathcal{Q}$ as functions of $\delta$ and $\Delta\mu_c$ (Fig.~\ref{fig:BennettSchematicSI}B-E).

%The error probability ($\eta$) and the energetic cost per step (affinity, $\mathcal{A}$) of this process are 
%determined by three key parameters: the binding free energies of correct and incorrect substrates (monomers), $\Delta\mu_c$ and $\Delta\mu_i$, and the difference in the kinetic barriers, $\delta$ (the parameters are specified in Fig.~\ref{fig:BennettSchematicSI}B). 

The evolution of the probability $P$ of the complementary polymer can be described by the following master equation,
\begin{align}
\dot{P}(\dots c) =& k^c_fP(\dots) + k^c_r P(\dots cc) + k^i_r P(\dots ci) \nonumber \\
                         &- (k^c_f+k^i_f + k^c_r)P(\dots c),  \nonumber \\
\dot{P}(\dots i)  =& k^i_fP(\dots) + k^c_r P(\dots ic) + k^i_r P(\dots ii) \nonumber\\ 
                         &- (k^c_f+k^i_f + k^i_r)P(\dots i),
\label{eqn:ME}
\end{align}
where the sequence of the complementary polymer is represented by ($\dots$), ($\dots c$), ($\dots i$), and so on as in Fig.~\ref{fig:BennettSchematicSI}A.

%Provided that the incorporation of correct and incorrect monomers have no correlation and is mutually independent,
%the average reaction currents of correct and incorrect monomer incorporation in the reaction scheme are given as 

Assuming that the error probability of each position of the copy polymer is independent from the prior sequence,
we can make the following substitutions: $P(\dots c) = P(\dots) P(c)$, $P(\dots ci) = P(\dots) P(c)P(i)$, and so forth.
Then, at steady state, Eq.~\ref{eqn:ME} takes the form
\begin{align}
 k^c_f + k^c_r P(c)^2 + k^i_r P(c)P(i) - (k^c_f+k^i_f + k^c_r)P(c)  &= 0 \nonumber \\
 k^i_f + k^c_r P(c)P(i) + k^i_r P(i)^2 - (k^c_f+k^i_f + k^i_r)P(i)  &= 0. \nonumber 
\end{align}
By rearranging the above equations, we can write the error probability, $\eta$, as
\begin{align}
\eta=P(i)=\frac{\langle J^i\rangle}{\langle J^i\rangle+\langle J^c\rangle}=\frac{k_f^i-\eta k_r^i}{k_f^i-\eta k_r^i+k^c_f - (1-\eta)k^c_r},
\label{eqn:eta}
\end{align}
where $\langle J^c\rangle$ and $\langle J^i \rangle$ are defined as the average reaction currents for correct and incorrect monomers, respectively,
\begin{align}
\langle J^c \rangle &= k^c_f - (1-\eta)k^c_r\nonumber\\
\langle J^i \rangle &= k^i_f - \eta k^i_r.
\end{align}
Essentially, we have transformed the dynamics along the tree structure (Fig~\ref{fig:BennettSchematicSI}A) 
into a Markov process. 
The equivalence of the expressions of $\eta$ from Eq.~\ref{eqn:ME} and Eq.~\ref{eqn:eta} has been shown for the double-cyclic reversible 3-state nework model in Ref.~\cite{Cady2009}. 
More general treatment of the dynamics that occur along tree structures can be found in refs.~\cite{Gaspard2014,Gaspard2016b,Gaspard2020}.
In the present work, the main conclusions pertaining to $\eta$ (and $\mathcal{Q}$) are further supported by explicit simulations of the master equation representation (Fig.~\ref{fig:PolPlot}C and Fig.~\ref{fig:Translation}D).

At the detailed balance (DB) condition, 
no current should flow through both pathways associated with correct and incorrect monomer incorporations,  
i.e., $\langle J^c\rangle =0$ and $\langle J^i\rangle = 0$; 
$\beta\mathcal{A}^c=\ln{\left(\frac{k_f^c}{(1-\eta)k_r^c}\right)}=0$ and $\beta\mathcal{A}^i=\ln{\left(\frac{k_f^i}{\eta k_r^i}\right)}=0$, 
which leads to 
\begin{align}
1-\eta_{eq}=\left(\frac{k_f^c}{k_r^c}\right)_{eq}\equiv e^{-\beta\Delta\mu_{c,eq}}
\label{eqn:DB1}
\end{align} 
and 
\begin{align}
\eta_{eq}=\left(\frac{k_f^i}{k_r^i}\right)_{eq}\equiv e^{-\beta\Delta\mu_{i,eq}}
\label{eqn:DB2}
\end{align} 
We note that in order for $\eta_{eq}$ to be in the range of $0<\eta_{eq}<1$, $\beta\Delta\mu_{c,eq}$ and $\beta\Delta\mu_{i,eq}$ should be positive, meaning that 
chemical potential bias of monomer incorporation is positive (uphill).  
Then, by taking the ratio between Eqs.~\ref{eqn:DB1} and \ref{eqn:DB2},
we obtain 
\begin{align}
\eta_{eq} = \frac{1}{1+e^{-\beta\Delta\Delta \mu_{ci}}}.
\label{eqn:defetaeq}
\end{align}
where $\Delta\Delta \mu_{ci}=\Delta\mu_{c,eq}-\Delta\mu_{i,eq}$.
At the limit of strongly forward driven reactions, i.e., $k_f^c\gg (1-\eta)k_r^c$ and $k_f^i\gg \eta k_r^i$, 
Eq.~\ref{eqn:eta} is led to 
\begin{align}
\eta \rightarrow k^i_f/(k^i_f+k^c_f) = 1/(1+e^{\beta\delta})\equiv \eta_0. 
\label{eqn:defeta0}
\end{align}
%With $\eta_0$ and $\eta_{eq}$, 
Next, we can write Eq.~\ref{eqn:eta} as
\begin{align}
\eta&=\frac{k_f^i/k_r^i-\eta}{k_f^i/k_r^i-\eta+(k_r^c/k_r^i)\left(k_f^c/k_r^c-(1-\eta)\right)}\nonumber\\
&=\frac{
e^{-\beta\Delta \mu_i}-\eta
}{
\left( e^{-\beta\Delta\mu_i}-\eta\right)+e^{-\beta(\Delta\Delta\mu_{ci}-\delta)}\left(e^{-\beta\Delta\mu_c}-(1-\eta)\right)
} 
\end{align}
where $k_r^c/k_r^i=e^{-\beta(\Delta\Delta\mu_{ci}-\delta)}$ was used. 
After some rearrangements, we can express $\Delta\mu_{c}$ as a function of $\eta$, $\eta_0$, and $\eta_{eq}$ as follows
\begin{align}
-\beta\Delta\mu_c &= \ln{ \frac{ \eta (1-\eta) (\eta_{eq}-\eta_{0}) }{( \eta-\eta_{0} ) \eta_{eq}} }.
\label{eqn:epsilonc}
\end{align}
Next, $\mathcal{A}$ can be written as
\begin{align}
\beta\mathcal{A}(\eta)&=\frac{1}{\langle J^c\rangle+\langle J^i\rangle}\left[\langle J^c\rangle\ln{\frac{k_f^c}{(1-\eta)k_r^c}}+\langle J^i\rangle \ln{\frac{k_f^i}{\eta k_r^i}}\right] \nonumber\\
&=(1-\eta)\ln{\frac{k_f^c}{(1-\eta)k_r^c}}+\eta\ln{\frac{k_f^i}{\eta k_r^i}}\nonumber\\
%&=-\beta\left[(1-\eta)\Delta\mu_c+\eta\Delta\mu_i\right]-[(1-\eta)\ln{(1-\eta)}+\eta\ln{\eta}]\nonumber\\
&= \underbrace{-\beta\left[(1-\eta)\Delta \mu_c+\eta \Delta\mu_i \right]}_{=-\beta\Delta\mu}+\underbrace{\left[-(1-\eta)\ln{(1-\eta)} - \eta\ln{\eta} \right]}_{=I}\nonumber\\
%&=-\beta\left[ \Delta\mu_c -\eta\Delta\Delta\mu_{ci}\right]-[(1-\eta)\ln{(1-\eta)}+\eta\ln{\eta}]\nonumber\\
%&=-\beta \Delta\mu_c +\eta\beta\Delta\Delta\mu_{ci}-[(1-\eta)\ln{(1-\eta)}+\eta\ln{\eta}]\nonumber\\
%&=-\beta \Delta\mu_c +\eta \ln { \frac{\eta_{eq}}{1-\eta_{eq}}}-[(1-\eta)\ln{(1-\eta)}+\eta\ln{\eta}]\nonumber\\
%&= \ln{ \frac{ \eta (1-\eta) (\eta_{0}-\eta_{eq}) }{( \eta-\eta_{0} ) \eta_{eq}} } +\eta \ln { \frac{\eta_{eq}}{1-\eta_{eq}}}-[(1-\eta)\ln{(1-\eta)}+\eta\ln{\eta}]\nonumber\\
%&=\eta \ln{\left[ \frac{\eta_{eq}(1-\eta)}{\eta (1-\eta_{eq}) }\right]}+\ln{\left[\frac{1-\eta_{eq}}{1-\eta}\right]}\nonumber\\
&= \eta \ln{ \frac{\eta_{eq}(1-\eta)}{\eta (1-\eta_{eq}) }} + \ln{\frac{\eta(\eta_{eq}-\eta_{0})}{\eta_{eq}(\eta-\eta_0)}}. 
\label{eqn:A}
\end{align}
As expected,  $\lim_{\eta\rightarrow\eta_{eq}}\beta \mathcal{A}(\eta)=0$ and $\lim_{\eta\rightarrow\eta_{0}}\beta \mathcal{A}(\eta)= \infty$, which means that $\eta$ approaches $\eta_{eq}$ and $\eta_{0}$ at the zero and infinite dissipation limits, respectively.

%At the DB condition ($\mathcal{A}=0$), the error probability ($\eta$) is determined by the difference in binding free energies as $\eta_{eq}[=1/(1+e^{-\beta{\Delta\Delta\mu_{ci}}})=f_0/(1+f_0)]$, where $\Delta\Delta\mu_{ci} = \Delta \mu_c-\Delta\mu_i$. 
%If the overall bias of the chemical potential is increased while keeping $\delta$ and $-\Delta\Delta\mu_{ci}$ constant, which amounts to increasing the overall substrate concentrations, the affinity is increased ($\mathcal{A} \gg 0$) and $\eta$ converges to $\eta_0=1/(1+e^{\beta\delta})$ (Fig.~\ref{fig:BennettSchematicSI}C). 
%When $\delta> -\Delta\Delta\mu_{ci}$ the enzyme can reduce the error probability from $\eta_{eq}$ to $\eta_0$
%by increasing $\mathcal{A}$.
%This error reducing strategy is referred to as \textit{kinetic discrimination}.

%For both $\lambda(\eta)$ and $\mathcal{A}(\eta)$, we used Eqs.~\ref{eqn:defetaeq},~\ref{eqn:defeta0}, and~\ref{eqn:epsilonc} to obtain the final expression.

Importantly, $\mathcal{A}$ can be decomposed into two contributions (Eq.~\ref{eqn:A}): $-\beta\Delta\mu$ is the free energy gain after the monomer incorporation, and $I$ is the Shannon information entropy arising from the chance of incorporating correct ($c$) and incorrect ($i$) monomers to the copy strand.  
The information ($I$) is
maximized to $I=\ln{2}$ when the odds of incorporating the correct and incorrect monomers is identical ($\eta=1/2$), whereas $I=0$ if only the correct or incorrect monomers are incorporated. 
This implies that as long as the chemical potential of monomers in solution is constantly maintained, the process near the DB condition ($\beta\mathcal{A}=-\beta\Delta\mu+I\gtrsim  0$) can still be driven by the entropy $I (\geq \beta\Delta\mu)$ even if the polymerization is energetically uphill ($\Delta\mu>0$) \cite{Bennett1976}. 
%Importantly, this fundamental property of copying enzymes also influences the behavior of TUR near equilibrium.
 
The Fano factor $\lambda$ can be calculated as 
\begin{align}
\lambda(\eta) &= 
\frac{\langle (\delta J^c)^2\rangle+\langle (\delta J^i)^2\rangle}{\langle J^c\rangle + \langle J^i \rangle}\nonumber\\ 
&= \frac{(k_f^i+\eta k_r^i)+(k^c_f + (1-\eta)k^c_r)}{(k_f^i-\eta k_r^i)+(k^c_f - (1-\eta)k^c_r)} \nonumber\\
&= \frac{
\left( k_f^i/k_r^i+\eta \right) +(k_r^c/k_r^i)\left(k_f^c/k_r^c+(1-\eta)\right)
}{
\left( k_f^i/k_r^i-\eta \right) +(k_r^c/k_r^i)\left(k_f^c/k_r^c-(1-\eta)\right)
} \nonumber\\
&=\frac{
\left( e^{-\beta\Delta\mu_i}+\eta\right)+e^{-\beta(\Delta\Delta\mu_{ci}-\delta)}\left(e^{-\beta\Delta\mu_c}+(1-\eta)\right)
}{
\left( e^{-\beta\Delta\mu_i}-\eta\right)+e^{-\beta(\Delta\Delta\mu_{ci}-\delta)}\left(e^{-\beta\Delta\mu_c}-(1-\eta)\right)
} \nonumber\\
%&=\frac{
%\left( \frac{\eta_{eq}}{1-\eta_{eq}}e^{-\beta\Delta\mu_c}+\eta\right)+\frac{1-\eta_0}{\eta_0}\frac{1-\eta_{eq}}{\eta_{eq}}\left(e^{-\beta\Delta\mu_c}+(1-\eta)\right)
%}{
%\left( \frac{\eta_{eq}}{1-\eta_{eq}}e^{-\beta\Delta\mu_c}-\eta\right)+\frac{1-\eta_0}{\eta_0}\frac{1-\eta_{eq}}{\eta_{eq}}\left(e^{-%\beta\Delta\mu_c}-(1-\eta)\right)
%} \nonumber\\
&=\frac{
2(\eta_{eq}-\eta_0)\eta^2 + (\eta_0+\eta_0^2-2\eta_{eq}) \eta +(1-\eta_0)\eta_0\eta_{eq}
}{
\eta_0(1-\eta_0)(\eta-\eta_{eq})
}.
\end{align}

$\mathcal{Q}(\eta)$ evaluated using the expression of $\mathcal{A}(\eta)$ and $\lambda(\eta)$, i.e., $\mathcal{Q}(\eta)=\mathcal{A}(\eta)\lambda(\eta)$, quantifies the translational efficiency of the copying enzyme along the template polymer \cite{Hwang2018JPCL,dechant2018JSM}.
For strongly driven systems $(\mathcal{A} \gg 0)$, all the curves of $\mathcal{Q}(\mathcal{A})$ with different values of $\delta$ converge (Fig.~\ref{fig:BennettSchematicSI}C). 
However, near the DB condition, where $I$ contributes significantly to $\mathcal{A}$, 
$\mathcal{Q}$ shows complex dependence on $\delta$. 
% and $\Delta\Delta\mu_{ci}$. 
%$\mathcal{Q}$ attains its lower bound  $2k_BT$ when $\delta = -\Delta\Delta\mu_{ci}$. %; otherwise $\beta\mathcal{Q}>2$.
%In particular, 
%For relatively large kinetic discriminations (i.e. $\delta \gg  -\Delta\Delta\mu_{ci}$), $\mathcal{Q} \gg 2~k_BT$.

%Notably, when $\delta \gg -\beta\Delta\Delta\mu_{ci}$, small energetic energetic drive ($\mathcal{A} \gtrsim 0$) can minimize $\eta$ and $\mathcal{Q}$ simultaneously (see magenta curve in Fig.~\ref{fig:BennettSchematic}C).

%TUR is also a function of $\eta$, $\mathcal{Q}(\eta)=\mathcal{A}(\eta)\lambda(\eta)$. 
%\begin{align}
%\mathcal{Q} &= \lambda \mathcal{A} \\
%&= \left[(\eta-\eta_{eq} )\lambda\right] \left[ \frac{\mathcal{A}}{\eta-\eta_{eq}} \right]
%\end{align}
%At the detailed balance condition, the numerator and denominator of the first term both converge to constants.
%The second term can be evaluated by differentiating the numerator and denominator by $\eta$. 
%Then, the final expression of $\mathcal{Q}$ at the detailed balance condition is
At the DB condition, 
\begin{align}
\beta \mathcal{Q}(\eta_{eq}) &= 2 + \frac{(\eta_{eq}-\eta_{0})^2}{(1-\eta_0)\eta_0} \geq 2.
\label{eqn:Qeq1}
\end{align}
Thus, the lower bound $2~k_BT$ is attained at the DB condition when $\eta_{eq}=\eta_{0}$.

$\mathcal{Q}$ can also approach its lower bound $2~k_BT$ at the limiting condition of $\beta\delta\gg 1$. 
At this limit, only correct monomers are incorporated into the copy polymer ($\langle J^i\rangle= 0$), which leads to
\begin{align}
\mathcal{Q} &= \mathcal{A}\frac{\langle (\delta J^c)^2 \rangle}{ \langle J^c \rangle} 
= \mathcal{A}\frac{ k^c_f + k^c_r }{ k^c_f - k^c_r } 
= \mathcal{A}\frac{ e^{\mathcal{\beta A}} + 1 }{ e^{\beta \mathcal{A}} - 1} \geq 2~k_BT,
\end{align}
and $\lim_{\mathcal{A}\rightarrow 0 , \delta\rightarrow\infty} \mathcal{Q} = 2~k_BT$.
The two limiting scenarios at which $\mathcal{Q}$ approaches $2~k_BT$ can be seen in Fig.~\ref{fig:BennettSchematicSI}E. 
%At the detailed balance condition,
%\begin{align}
%\lim_{\mathcal{A}\rightarrow 0 , \delta\rightarrow\infty} \mathcal{Q} = 2~k_BT.
%\label{eqn:Qeq2}
%\end{align}
%At first glance, Eqs.~\ref{eqn:Qeq1} and~\ref{eqn:Qeq2} may seem to contradict each other.
%When $\delta\rightarrow\infty$, $\eta_0\rightarrow0$, and $\mathcal{Q}$ of Eq.~\ref{eqn:Qeq1} diverges to infinity.
%However, Eq.~\ref{eqn:Qeq1} is defined only for strictly positive values of $\eta$, as can be seen in the expressions of $\lambda$ and $\mathcal{A}$ ( Eqs.~\ref{eqn:defetaeq} and~\ref{eqn:defeta0}). In sum, the equilibrium $\mathcal{Q}$ of copy processes is a discontinuous function of $\delta$.

\begin{figure*}[t]
	\begin{center}
	\includegraphics[width=0.95 \textwidth]{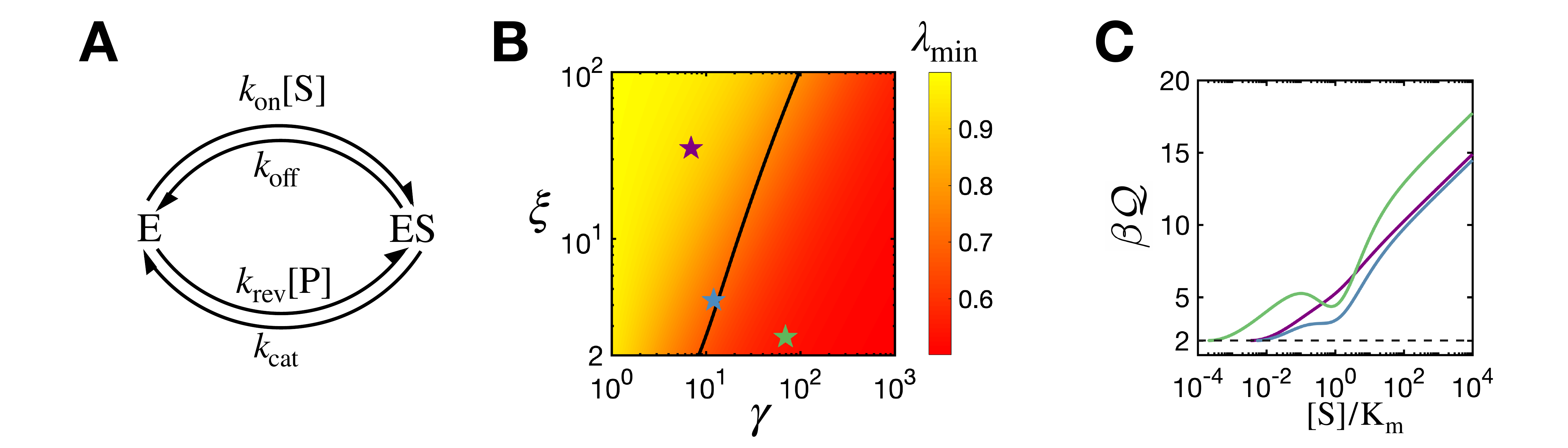}
	\caption{
	Michaelis-menten type enzyme kinetics. 
	{(A)} Schematic of MM type enzyme kinetics. 
	{(B)} The minimum Fano factor with respect to $\text{[S]}$ ($\lambda_{\rm min})$ plotted against dimensionless constants
	$\gamma = \frac{k_{\text{cat}}}{ \sqrt{k_{\text{off}}k_{\rm rev}[\text{P}]}}$ and
	$\xi =  \frac{k_{\text{off}}+k_{\rm rev}[\text{P}]}{ \sqrt{k_{\text{off}}k_{\rm rev}[\text{P}]}}$. 
	To the left (resp. right) of the black curve, $\mathcal{Q}$ is monotonic (resp. non-monotonic).
	{(C)} $\mathcal{Q}$ plotted as a function of $\text{[S]}$ normalized by ${K_m}=\frac{k_{\text{off}}+k_{\text{cat}}}{k_{\text{on}}}$. 
	The corresponding $\gamma$ and $\xi$ values are shown in (B) by the star symbols with the matching color. 
	The dotted line demarcates the lower bound $\mathcal{Q}=2k_BT$. 
	The kinetic rate constants used for the plots are as follows:
	Green: $k_{\text{on}} = 10^8 $ $ \text{M}^{-1}{s}^{-1}$, $k_{\text{off}} = 30$ $s^{-1}$, $k_{\text{cat}} =10^3$ $s^{-1}$, $k_{\text{rev}}[\text{P}] =6.9$ $s^{-1}$;
	Blue: $k_{\text{on}} = 10^8 $ $ \text{M}^{-1}{s}^{-1}$, $k_{\text{off}} = 10$ $s^{-1}$, $k_{\text{cat}} =30$ $s^{-1}$, $k_{\text{rev}}[\text{P}] =6.2\times10^{-1}$ $s^{-1}$;
	Purple: $k_{\text{on}} = 10^8 $ $ \text{M}^{-1}{s}^{-1}$, $k_{\text{off}} = 50$ $s^{-1}$, $k_{\text{cat}} =10$ $s^{-1}$, $k_{\text{rev}}[\text{P}] =4.1\times10^{-2}$ $s^{-1}$.	
	}
	\label{fig:TwoState}
	\end{center}
\end{figure*}

\subsection*{The suboptimal condition of reversible Michaelis-Menten reactions}
We provide conditions at which $\mathcal{Q}$ has a local minimum with respect to substrate concentration ($[S]$) in reversible MM reactions shown in Fig.~\ref{fig:TwoState}A. 
First, the affinity ($\mathcal{A}$),
\begin{align}
\beta\mathcal{A}([S]) = \ln{\left(\frac{k_{\text{on}}k_{ \text{cat} }[S]}{k_{\text{off}}k_{\text{rev}}[P]}\right)},
\end{align}
is a strictly increasing function of $\text{[S]}$. 
Next, the Fano factor ($\lambda$) as a function of $\text{[S]}$ is given by 
\begin{align} 
\lambda([S]) &= \frac{  
k_{\text{on}}k_{\text{cat}}[S] + k_{\text{off}}k_{\text{rev}}[P]
-2 \left( \frac{ k_{\text{on}}k_{\text{cat}}[S] - k_{\text{off}}k_{\text{rev}}[P] }{ k_{\text{on}}[S] + k_{\text{off}} + k_{\text{cat}} + k_{\text{rev}}[P] } \right)^2
  }{k_{\text{on}}k_{\text{cat}}[S] - k_{\text{off}}k_{\text{rev}}[P] } \\
 &=  \frac{e^{\beta \mathcal{A}}+1}{e^{\beta \mathcal{A}}-1} 
- 2\frac{ \gamma^2 \left(  e^{\beta \mathcal{A}}-1 \right) }{ \left( \gamma^2+\gamma\xi +e^{\beta \mathcal{A}} \right)^2 } ,
\end{align}
where $\gamma = \frac{k_{\text{cat}}}{\sqrt{{k_{\text{off}}}{k_{\rm rev}}[P]}}$ and $\xi = \frac{k_{\text{off}}+k_{\rm rev}[P]}{\sqrt{k_{\text{off}}k_{\rm rev}[P]}}$ are dimensionless constants. 
%Note, $\gamma>0$ and $\xi \geq 2$ by construction.
%The conditions at which $\lambda([\text{S}])$ and  $\lambda(\mathcal{A})$ have critical points are identical, since $ \frac{\partial \lambda}{{\partial {[\text{S}]} }} =\frac{ \partial \lambda}{{\partial A} }\frac{\partial \mathcal{A}}{{\partial {[\text{S}]} } }$, and $\frac{\partial \mathcal{A}}{{\partial {[\text{S}]} }} >0$ for all $[\text{S}]$.
%The condition $\frac{ \partial \lambda}{{\partial A} }=0$ leads to a 3rd degree polynomial of $e^{\mathcal{\beta A}}$, which has real roots only when $\gamma>1$. 
$\lambda$ has a local minimum only when $\gamma>1$, or equivalently, when $k_{\text{cat}} > \sqrt{k_{\text{off}}k_{\rm rev}[P] }$. 
%Solving for $\frac{ \partial \lambda}{{\partial A} }=0$ leads to a 3rd degree polynomial of $e^{\mathcal{\beta A}}$, which has one real root ($\mathcal{A}_{cr}$), with the following expression
%\begin{align}
%\beta \mathcal{A}_{cr} = 
%\ln{\left[\frac{1}{3(\gamma^2-1)} 
%\left(
%\frac{X}{Y}+Y+Z
%\right) \right]},
%\label{eqn:Acr}
%\end{align}
%where 
%\begin{align}
%X =& \gamma^2(15+\gamma^2)\left(1+\gamma(\gamma+\xi)\right)^2,\\
%Y = & \bigg[\gamma^2(27+36\gamma^2+ \gamma^4) + \left(1+\gamma(\gamma+\xi)\right)^3 \\
%     &+ 3\sqrt{3} \sqrt{\gamma^4 (\gamma^2-1)^2(27+\gamma^2)\left( 1+\gamma(\gamma+\xi) \right)^6} \bigg]^{1/3},\\
%Z =& \gamma \left( 3 \xi + \gamma (7 + \gamma(\gamma+\xi)) \right).
%\end{align}
%After some rearrangements, we can show that $Z-{3(\gamma^2-1)}=\left(3+\gamma^2\right)\left(1+\gamma(\gamma+\xi)\right)$.
%Thus, $\frac{Z}{{3(\gamma^2-1)}}>1$ as long as $\gamma>1$.
%Since $X,Y>0$ as well, $\mathcal{A}_{cr}>0$  if and only if $\gamma>1$. 
%After some rearrangements, it can be shown that $\beta \mathcal{A}_{cr}>0$  if and only if $\gamma>1$. 
%Thus, $\lambda[\text{S}]$ has a local minimum only when $\gamma>1$, or equivalently, when $k_{\text{cat}} > \sqrt{k_{\text{off}}k_{\rm rev}}$. 
At the limit of a strongly driven catalytic step  ($k_{\text{cat}}\gg k_{\text{rev}}[P]$), the expression for $\lambda$ simplifies to 
\begin{align}
\lambda([S])\approx 1-\frac{2k_{\text{cat}}[S]}{k_{\text{on}}([S]+K_\text{m})^2},
\end{align} 
which is minimized at $[S]={ K_m}$ with ${ K_m}=(k_{\text{off}}+k_{\text{cat}})/k_{\text{on}}$ (Fig.~\ref{fig:TwoState}B). 

Since $\mathcal{A}$ is monotonic with $\text{[S]}$, a local minimum of $\mathcal{Q}(\text{[S]})=\mathcal{A}(\text{[S]})\lambda(\text{[S]})$ can only occur near to that of $\lambda(\text{[S]})$. 
Using this, we numerically determined the range of $\gamma$ and $\xi$ values
at which $\mathcal{Q}(\text{[S]})$ has a local minimum away from the DB condition. 
%To determine whether or not $\mathcal{Q}(\text{[S]})$ is a monotonically increasing function, 
%we computed $\frac{\partial \mathcal{Q}}{\partial \mathcal{A}}$ for the range of values $ 0 < \mathcal{A} < 10 \mathcal{A}_{cr}$,
%where $\mathcal{A}_{cr}$ is the critical point of $\lambda$. 
For any $\xi$, 
there exists a $\gamma$ above which $\mathcal{Q}\text{[S]}$ is non-monotonic (Fig.~\ref{fig:TwoState}B). 
Thus, when $k_{\text{cat}} \gg k_{\rm rev}[P]$ and $k_{\text{cat}}$ is sufficiently larger than $\sqrt{k_{\text{off}}k_{\rm rev}[P]}$,
$\mathcal{Q}(\text{[S]})$ has a local minimum around $[\text{S}] \approx { K_m}$.

\subsection*{Mathematical expressions of $\eta$ and $\mathcal{A}$ for general copying processes}
Here, we provide the details of obtaining $\eta$ and $\mathcal{Q}$ in the main text. A more mathematically rigorous treatment on the subject can be found in Ref. \cite{Pigolotti2016}.

%Consider a copying enzyme at the $l^{\text{th}}$ position of a template polymer, at which $M$ types of monomers can be incorporated into the copy %polymer.
%For each monomer of the $m^{\text{th}}$ type, where $m\in \{1,\dots,M\}$, there are $P_m$ different kinetic pathways by which the monomer can be %incorporated. 
%For mRNA translation, $M=20$ represents the number of all amino-acid types.
%If the codon at position $l$ of the mRNA is GGG, the amino-acids Gly, Ala, Arg, Glu, Trp, and Val can be incorporated into the protein, with the following %number of different aa-tRNAs available for each: $P^l_{\rm Gly}=3,~P^l_{\rm Ala}=1,~P^l_{\rm Arg}=2,~P^l_{\rm Glu}=1,~P^l_{\rm Trp}=1,$ and $P^l_{\rm %Val}=1$ (Fig.~\ref{fig:codon_table}).

In the following, without loss of generality, we will define the error probability ($\eta$) and affinity ($\mathcal{A}$) of copy processes by referring to the kinetic mechanism of mRNA translation.
To begin, consider the ribosome at position $l$ of the mRNA sequence,
decoding a specific codon type.
For each amino-acid type $a$, there exists a set $T^l_a$, of associated aa-tRNAs, 
each of which represents a separate kinetic path of incorporating $a$ into the protein. 
If the codon at position $l$ is GGG, amino-acids Gly, Ala, Arg, Glu, Trp, and Val can be polymerized, with the following set of associated tRNAs: $T^l_{\rm Gly}=\{  \rm tRNA^{Gly}_{1} ,~ tRNA^{Gly}_{2},~ tRNA^{Gly}_{3} \}$, $T^l_{\rm Ala}=\{ \rm tRNA^{Ala}_{1B}\}$, $T^l_{\rm Arg}=\{ \rm tRNA^{Arg}_{3},~ tRNA^{Arg}_{5}\}$, $T^l_{\rm Glu}= \rm \{ tRNA^{Glu}_{2}\}$, $T^l_{\rm Trp}= \rm \{ tRNA^{Trp}_{}\},$ and $ T^l_{\rm Val}= \rm \{ tRNA^{Val}_{1}\}$ (Fig.~\ref{fig:codon_table}).

At steady state, we assume that, $\eta^l_{a}$, the probability of incorporating amino-acid $a$ at position $l$, where the codon is specified, is given by 
\begin{align}
\eta^l_a = \frac{\sum_{t \in  T^l_a}\langle J^{l,\text{pol}}_{a,t} \rangle }{ \sum_{\alpha \in \{\text{aa}\}}\sum_{t \in T^l_{\alpha}} \langle J^{l,\text{pol}}_{\alpha,t}  \rangle }, 
\label{eqn:generaleta}
\end{align}
where $\{\text{aa}\}$ is the set of all amino-acids,
and $\langle J^{l,\text{pol}}_{a,t} \rangle $ is the polymerization current of 
aa-tRNA $t$.

Next, we define the affinity associated with polymerization 
using the previously discussed example of Gly incorporation at codon GGG.
The polymerization affinity of Gly along the cognate kinetic path associated with $\rm Gly\text{-}tRNA^{Gly}_{1}$ is
\begin{align}
\beta\mathcal{A}^{l,\text{pol}}_{\rm Gly,\rm Gly1} 
%&= \ln{ \left(\frac{k_{\text{on}}[{\rm Gly\text{-}tRNA^{Gly}_{1}}]k_{ \text{rec},f}k^{\text{C}}_{ \text{hyd},f}k^{\text{C}}_{\text{pol}}}
%{\eta^l_{\rm Gly} k_{\text{off}}k^{\text{C}}_{ \text{rec},r}k^{\text{C}}_{ \text{hyd},r}k^{\text{C}}_{\text{dep}}}\right) } \\
&= -\beta\Delta\mu^{l,\text{pol}}_{\rm Gly,Gly1} - \ln{(\eta^l_{\rm Gly})},
\end{align}
where $-\beta\Delta\mu^{l,\text{pol}}_{\rm Gly,Gly1}=\ln{ \left(\frac{k_{\text{on}}[\text{C}]k_{ \text{rec},f}k^{\text{C}}_{ \text{hyd},f}k^{\text{C}}_{\text{pol}}}
{k_{\text{off}}k^{\text{C}}_{ \text{rec},r}k^{\text{C}}_{ \text{hyd},r}k^{\text{C}}_{\text{dep}}}\right) }$,
and [C] represents the concentration of the ternary complex $({\rm Gly\text{-}tRNA^{Gly}_{1})\text{-}(EF\text{-}Tu)\text{-}GTP}$ (Fig.~\ref{fig:Ribosome}B).
The term $\ln{(\eta^l_{\rm Gly})}$ is required to account for the fact that 
Gly can be depolymerized at position $l$ only $\eta^l_{\rm Gly}$ fraction of the time. 
Similarly, the affinity of incorporating Gly along the near-cognate kinetic path associated with $\rm Gly\text{-}tRNA^{Gly}_{3}$ is
\begin{align}
\beta\mathcal{A}^{l,\text{pol}}_{\rm Gly,\rm Gly3} 
&= -\beta\Delta\mu^{l,\text{pol}}_{\rm Gly,Gly3} - \ln{( \eta_{\rm Gly}^l)},
\end{align}
where $-\beta\Delta\mu^{l,\text{pol}}_{\rm Gly,Gly3}=\ln{ \left(\frac{k_{\text{on}}[\text{NC}]k_{ \text{rec},f}k^{\text{NC}}_{ \text{hyd},f}k^{\text{NC}}_{\text{pol}}}
{ k_{\text{off}}k^{\text{NC}}_{ \text{rec},r}k^{\text{NC}}_{ \text{hyd},r}k^{\text{NC}}_{\text{dep}}}\right) }$,
and $[\text{NC}]=[({\rm Gly\text{-}tRNA^{Gly}_{3}})\text{-}(\rm EF\text{-}Tu)\text{-}GTP]$.
Generally, we denote the polymerization affinity of amino-acid $a$ associated with aa-tRNA $t$ by $\mathcal{A}^{l,\text{pol}}_{a,t}$.

%To define the affinity of the copying process, we begin by defining $(1)$, $(2)$, $\dots$, $(x)$ to be a sequence of intermediate states in the $p^{\rm th}$ polymerization pathway of the $m^{\rm th}$ monomer (e.g., $x=3$ for a 3-state kinetic cycle). Next, let $\kappa^{1,2}_{p,m}$, \dots,$\kappa^{x,x+1}_{p,m}$ and $\kappa^{2,1}_{p,m}$, \dots,$\kappa^{x+1,x}_{p,m}$ be the respective forward and reverse rate constants for the reactions defined along the sequence of states $(1)$, $(2)$, $\dots$, $(x)$, $(1)$. 
%Then, the polymerization affinity of the $m^{\rm th}$ monomer along the $p$-th kinetic pathway is 
%\begin{align}
%\beta\mathcal{A}^{p,m}_{\text{pol}} =\ln{\left( \frac{\prod_{l=1}^{x} \kappa^{l,l+1}_{p,m}}{\eta_{m}\prod_{l=1}^{x} \kappa^{l+1,l}_{p,m}}\right)}= -\beta\Delta\mu_{\text{pol}}^{p,m} + \ln{\left(\frac{1}{\eta_m}\right)},
%\end{align}
%where $-\beta\Delta\mu^{p,m}_{\text{pol}}=\ln{\left( \frac{\prod_{l=1}^{x} \kappa^{l,l+1}_{p,m}}{\prod_{l=1}^{x} \kappa^{l+1,l}_{p,m}}\right)}$. 
%The term $ \ln{\left(\frac{1}{\eta_m}\right)}$ is required to account for the fact that only $\eta_m$ fraction of state (1) can depolymerize the $m^{\text{th}}$ monomer.
%Note that the polymerization affinity of monomer $m$ can vary depending on the kinetic pathway $p$. 
%For instance, a single amino-acid can be associated with multiple (aa-tRNA)-(EF-Tu)-GTP complexes present at different concentrations in cellular milieu.

Next, we let $\langle J_{a,t}^{l,\text{fut}}\rangle$ and $\Delta\mu_{a,t}^{l,\text{fut}}$ be the current and affinity of the futile cycle 
within the incorporation path of aa-tRNA $t$ associated with amino-acid $a$.
Denoting the net polymerization flux by $\langle J^{l}\rangle\left(=\sum_{a \in\{aa\}}\sum_{t \in T_a^l} \langle J^{l,\text{pol}}_{a,t} \rangle\right)$,
we can write the affinity of mRNA translation at position $l$ as
\begin{align}
\beta\mathcal{A}^l=& \underbrace{
{-\beta}
 \sum_{a \in \{\text{aa}\}} \sum_{ t \in T_a^l}
\left( 
 \frac{\langle J^{l,\text{pol}}_{a,t} \rangle}{\langle J^l \rangle} 
\Delta\mu^{l,\text{pol}}_{a,t}  + 
\frac{\langle J^{l,\text{fut}}_{a,t} \rangle}{\langle J^l \rangle} \Delta\mu^{l,\text{fut}}_{a,t} \right)
}_{=-\beta\Delta\mu} \underbrace{-\sum_{a\in\{\text{aa}\}} \eta^l_a \ln \eta^l_a }_{=I}.
\label{eqn:generalA}
\end{align}
For the Bennett model, which involves two types of monomers, with one incorporation pathway each, without any futile cycles, we recover Eq.~\ref{eqn:A}.
To estimate $I$ in the extended model of translation (Fig.~\ref{fig:Translation}), we sum the Shannon-entropy of each position $l$ along the protein sequence. 
% (Eq.~\ref{eqn:extendedI}).

%Let $\langle J_{p,m,f}\rangle$ and $\mu_{\rm fut}^{p,m,f}$ be the current and affinity of the $f^{\text{th}}$ futile cycle within 
%the corresponding incorporation process. For instance, in the translation model, there is one futile cycle in all kinetic pathways of all types of amino-acids.
%Denoting the net polymerization flux by $\langle J\rangle\left(=\sum_{p=1}^{P_m}\sum_{m=1}^{M} \langle J_{p,m} \rangle\right)$,
%we can write the affinity of the copying process as
%\begin{align}
%\beta\mathcal{A}=& \underbrace{
%{-\beta}
%\sum_{p=1}^{P_m} \sum_{m=1}^{M}
%\left( 
% \frac{\langle J_{p,m} \rangle}{\langle J \rangle} 
%\Delta\mu_{\text{pol}}^{p,m}  +
%\sum_{f} \frac{\langle J_{p,m,f} \rangle}{\langle J \rangle} \Delta\mu^{p,m,f}_{\rm fut} \right)
%}_{=-\beta\Delta\mu} \nonumber\\
%&\underbrace{-\sum_{m=1}^{M} \eta_m \ln \eta_m }_{=I}, 
%\label{eqn:generalA}
%\end{align}
%where the index $f$ runs through all the futile cycles for each kinetic pathway. 
%The affinity of the futile cycle, $\Delta\mu_{p,m,f}$, is independent of $\eta_m$, since futile cycles initiate and terminate at the same states.

\subsection*{Computation of $\eta$ and $\mathcal{Q}$}
We will work through the process of calculating $\eta$ and $\mathcal{Q}$ in the T7 DNA polymerase model, 
for which we apply Koza's method of calculating currents and fluctuations in kinetic networks \cite{Koza1999JPA}. 
To simplify the notation, we will relabel the states in Fig.~\ref{fig:PolPlot}A by indices 1 through 5; 
i.e., $\text{E}^{(1)} \rightarrow 1$, $c^{(2)} \rightarrow 2$, $c^{(3)} \rightarrow 3$, $i^{(2)} \rightarrow 4$, and $i^{(3)} \rightarrow 5$.
Additionally, we will relabel the rate constants so that $k_{\mu,\nu}$ represents the rate of the reaction from state $\mu$ to $\nu$;
i.e., $k_{1,2} = k^c_{\text{on}} [\text{dNTP}]$, $k_{2,3} = k^c_{\text{conf}}$, and so forth.
Here, the depolymerization rate constants are set to 
$k_{1,3} = (1- \eta) k^c_{\text{dep}}$ and $k_{1,5} = \eta k^i_{\text{dep}}$.

%Then, we can write the evolution of the probability of states $P(\mu)$ for $\mu=1,\dots,5$ as
%\begin{align}
%\frac{ dP(\mu)}{dt} = \sum_{\nu=1}^{5}[\Gamma]_{\mu \nu} P(\nu),
%\end{align}
%where $\Gamma$ is a matrix shown in Eq.~\ref{eqn:Gamma}.
%To will compute the current of correct nucleotide incorporation $ \langle J^c_{\text{pol}} \rangle $,
%we construct matrix $\Gamma^{c}(z)$ as in Eq.~\ref{eqn:Gammac}.
 % $\Gamma^{c}(z)$  is a slightly modified version of  $\Gamma$, in which we multiply $e^z$ and $e^{-z}$ to the rates $k^c_{\rm po}$ and $k^c_{\text{dep}$, respectively.
 %Next, let $C_n(z)$ denote the coefficients of the characteristic polynomial of $\Gamma^c(z)$ (i.e., $\sum_{n=0}^{5} C_n(z) \Lambda^c_0(z)^n = 0$). 
 %Then, it can be shown that the following expressions hold for $\langle J^c_{\text{pol}} \rangle$ and $\langle (\delta J^c_{\text{pol}})^2\rangle$,
%\begin{align}
%\langle J^c_{\text{pol}} \rangle &= \left(\Lambda^c_0\right)'= -\frac{C'_0}{C_1}, \\
%\langle (\delta J^c_{\text{pol}})^2 \rangle &= \left(\Lambda^c_0\right)''= -\frac{C''_0 + 2C'_1(\Lambda^c_0)'+2C_2(\Lambda^c_0)'}{C_1}
%\end{align}
%where $\prime$ denotes the derivative with respect to $z$ evaluated at $z=0$. 

To begin, we will compute the current of correct nucleotide incorporation $ \langle J^c_{\text{pol}} \rangle $.
For the $i$-th chemical state ($i \in \{ 1,2, \dots,N\}$, where $N=5$ ) at time $t$,
let $\mu(t) \equiv i + N\times n^c(t) $ be the generalized state of the system after completing $n^c(t)$ correct nucleotide incorporation cycles. 
Then, let $P(\mu,t)$ be the probability of the system to be in state $\mu$ at time $t$. The time evolution of $P(\mu,t)$ is given by
\begin{align}
\frac{\partial P(\mu,t) }{\partial t} &= \sum_{\xi} \left [ k_{\mu-\xi,\mu} P(\mu-\xi,t) 
-  k_{\mu,\mu-\xi} P(\mu,t) \right],
\label{eqn:mastereq}
\end{align}
where the index $\xi$ runs through all states one reaction away from state $\mu$. 
Here, the periodicity of the network model constrains the rate constants so that $k_{\mu,\nu} = k_{i,j}$ for $\mu = i \pmod{N}$ and $\nu=j\pmod{N}$.
Following Ref.~\cite{Koza1999JPA}, we define $P_j(\mu,t)$ as
\begin{align}
P_j(\mu,t) &\equiv P(\mu,t)\delta^N_{\mu,j},
\end{align}
where 
\begin{equation}
  \delta^N_{\mu,j} =
    \begin{cases}
      1, & \text{if $j=\mu \pmod N$} \nonumber\\
      0, & \text{otherwise,}
    \end{cases}       
\end{equation}
for $j \in \{1,2,\dots,N\}$. 
By multiplying $\delta^N_{\mu,j}$ to both sides of Eq.~\ref{eqn:mastereq}
and using the equality $\delta^N_{\mu,j} = \delta^N_{\mu-\xi,j-\xi}$, we get
\begin{align}
\frac{\partial P_j(\mu,t) }{\partial t} &= \sum_{\xi} \left [ k_{j-\xi,j} P_{j-\xi}(\mu-\xi,t) 
-  k_{j,j-\xi} P_j(\mu,t) \right].
\end{align}

To derive the expression of $\langle J^c_{\text{pol}} \rangle$ as a function of the rate constants, we define
the generating function 
\begin{align}
\mathcal{G}^{c}_{j}(z,t) &\equiv \sum_{\mu = -\infty}^{\infty} e^{z X_{\mu}} P_j(\mu,t),
\end{align}
where $X_{\mu}$ is the coordinate for the correct nucleotide incorporation cycle at state $\mu$.
Then, the time derivative of the generating function can be written as 
%\begin{align}
%\frac{ \partial \mathcal{G}^{c}_{j}(z,t) }{\partial t} &= 
% \sum_{\xi}  e^{z d_{j-\xi,j}}k_{j-\xi,j} \mathcal{G}^{c}_{j-\xi}(z,t)  
%-\sum_{\xi} k_{j,j-\xi} \mathcal{G}^{c}_{j}(z,t),  
%\label{eqn:generator1}
%\end{align}
%where 
%\begin{equation}
 % d_{i,j} =
  %  \begin{cases}
  %    1, & \text{if $i=3$ and $j=1$}\\
  %   -1, & \text{if $i=1$ and $j=3$}\\
  %    0, & \text{otherwise}.
  %  \end{cases}       
%\end{equation}
%Eq.~\ref{eqn:generator1} can be expressed more succinctly as 
\begin{align}
\frac{\partial \mathcal{G}^{c}_{j}(z,t)}{\partial t} &= \sum_{i=1}^{N}\Gamma^{c}_{i,j}(z) \mathcal{G}^c_i(z,t),
\end{align}
where the matrix $\Gamma^{c}(z)$ is defined as 
\begin{equation}
\Gamma^{c}_{i,j}(z) = 
    \begin{cases}
	k_{i,j} e^{z d_{i,j}}, & \text{if $i\neq j$}\nonumber\\
	-\sum_{m=1(\neq i)}^{N} k_{i,m}, & \text{if $i=j$},
    \end{cases}       
\end{equation}
and also shown in the matrix form below
\begin{widetext}
\scriptsize
\begin{align}
\begin{bmatrix} 
-\left( k_{\text{on}}^c \text{[dNTP]} + k_{\text{on}}^i \text{[dNTP]} + k_{\text{dep}}^c (1-\eta) + k_{\text{dep}}^i \eta \right) 
& k_{\text{on}}^c\text{[dNTP]} & k_{\text{dep}}^c(1-\eta)e^{-z} & k_{\text{on}}^i\text{[dNTP]} & k_{\text{dep}}^i \eta \\
k_{\text{off}}^c & -\left( k_{\text{off}}^c + k_{\text{conf},f}^c   \right) & k_{\text{conf},f}^c  & 0 & 0 \\
k_{\text{pol}}^c e^{z}& k_{\text{conf},r}^c  & -\left( k_{\text{pol}}^c + k_{\text{conf},r}^c \right) & 0 & 0 \\
k_{\text{off}}^i & 0& 0& -\left( k_{\text{off}}^i + k_{\text{conf},f}^i   \right) & k_{\text{conf},f}^i \\
k_{\text{pol}}^i & 0& 0& k_{\text{conf},r}^i & -\left( k_{\text{pol}}^i+ k_{\text{conf},r}^i \right)
\end{bmatrix}
\label{eqn:Lcmatrix}
\end{align}
\normalsize
\end{widetext}

For the computation of $\langle J^{c}_{\rm pol}\rangle$, $d_{i,j}$ is defined as 
\begin{equation}
  d_{i,j} =
   \begin{cases}
     1, & \text{if $i=3$ and $j=1$}\\
    -1, & \text{if $i=1$ and $j=3$}\\
     0, & \text{otherwise}.
   \end{cases}       
\end{equation}

Next, we define 
$\mathcal{G}^{c}(z,t) \equiv \sum_{i=1}^{N} \mathcal{G}^{c}_i(z,t)$
and denote the coordinate of the correct incorporation cycle by $X^c(t)$.
%\begin{align}
%\mathcal{G}^{c}(z,t) \equiv \sum_{i=1}^{N} \mathcal{G}^{c}_i(z,t) = e^{\Lambda^{c}_0(z) t,
%\end{align}
%where $\Lambda^{c}_0(z)$ is the maximum eigenvalue of the matrix $\Gamma^{c}(z)$. 
Then, it can be shown that 
\begin{align}
\langle J^c_{\text{pol}} \rangle &= 
\lim_{t \rightarrow \infty} \frac{ \langle X^c(t) \rangle }{t} 
= \lim_{t \rightarrow \infty} \frac{\partial_z\mathcal{G}^c(z,t)\vert_{z=0}}{t}
= \partial_z\Lambda^c_0(z)\vert_{z=0},  
\end{align}
and 
\begin{align}
\big\langle \left( \delta  J^c_{\text{pol}} \right)^2 \big\rangle
&= \lim_{t\rightarrow \infty} \frac{\langle (X^c(t))^2 \rangle - \langle X^c(t)\rangle^2}{t} \nonumber\\
&= \lim_{t\rightarrow \infty} \frac{  \partial_z^2\mathcal{G}^c(z,t)\vert_{z=0}- \left(\partial_z\mathcal{G}^c(z,t)\vert_{z=0} \right)^2}{t} \nonumber\\
&= 
\partial_z^2 \Lambda_0^c (z)\vert_{z=0},
\end{align}
where $\Lambda^{c}_0(z)$ denotes the maximum eigenvalue of the matrix $\Gamma^{c}(z)$.
%where the apostrophe denotes the partial derivative with respect to $z$ evaluated at $z=0$. 
%The partial derivatives of $\mathcal{G}^c(z,t)$ are also all evaluated at $z=0$.
Now, let $C_n(z)$ denote the coefficients of the characteristic polynomial of $\Gamma^c(z)$ 
(i.e., $\sum_{n=0}^{N} C_n(z) \Lambda^c_0(z)^n = 0$). 
Then, we can write the following expressions for $\langle J^c_{\text{pol}} \rangle$ and $\langle (\delta J^c_{\text{pol}})^2\rangle$,
\begin{align}
\langle J^c_{\text{pol}} \rangle &= \left(\Lambda^c_0\right)'= -\frac{C'_0}{C_1}, \\
\langle (\delta J^c_{\text{pol}})^2 \rangle &= \left(\Lambda^c_0\right)''= -\frac{C''_0 + 2C'_1(\Lambda^c_0)'+2C_2\left((\Lambda^c_0)'\right)^2}{C_1}.
\end{align}
where $\prime$ denotes the derivative with respect to $z$ evaluated at $z=0$, and $C_1$ and $C_2$ are evaluated at $z=0$.
We can analogously compute the current of incorrect monomer incorporation, $\langle J^i_{\text{pol}} \rangle $, by constructing the corresponding matrix $\Gamma^i(z)$, in which the non-diagonal entries of $k_{5,1}$ and $k_{1,5}$ are multiplied by $e^{z}$ and $e^{-z}$, respectively. 
Since $k_{1,3}$ and $k_{1,5}$ are functions of $\eta$, $\langle J^c_{\text{pol}} \rangle $ and $\langle J^i_{\text{pol}} \rangle $ are functions of $\eta$. Thus, we can solve for $\eta$ by the equality
\begin{align}
\eta = \frac{\langle J^i_{\text{pol}} \rangle}{\langle J^i_{\text{pol}}\rangle + \langle J^c_{\text{pol}} \rangle}.
\end{align}
After we obtain the numerical expression of $\eta$, we can construct the matrix $\Gamma(z)$ to calculate the total flux $\langle J_{\text{pol}} \rangle=\langle J^c_{\text{pol}} \rangle+\langle J^i_{\text{pol}} \rangle$ and its fluctuation $\langle  ( \delta J_{\text{pol}})^2 \rangle$. 
With known values of $\eta$ and $\langle J_{\text{pol}}\rangle$, the affinity of replication ($\mathcal{A}$ ) can be computed by Eq.~\ref{eqn:generalA}. 
In sum, we have demonstrated how to calculate $\mathcal{Q}=\mathcal{A}\langle (\delta J_{\text{pol}})^2\rangle/\langle J_{\text{pol}}\rangle$ for the T7 DNA polymerase model.

We can calculate $\eta$ and $\mathcal{Q}$ of the simplified ribosome model in a similar way. 
To obtain $\langle J^{\text{C}}_{\text{pol}} \rangle$ as a function of $\eta$, we construct the corresponding matrix $\Gamma^{\text{C}}(z)$, 
in which we multiply the non-diagonal entries corresponding to $k_{\text{pol},f}^{\text{C}}$ and $k_{\text{pol},r}^{\text{C}}$ by $e^z$ and $e^{-z}$, respectively.
The expression for $\langle J^{\text{NC}}_{\text{pol}} \rangle$ is obtained analogously, by constructing the corresponding matrix $\Gamma^{\text{NC}}(z)$. 
After solving for the numerical value of the error probability, we can calculate the total polymerization rate $\langle J_{\text{pol}} \rangle$ 
and its fluctuation  $\langle (\delta J_{\text{pol}})^2 \rangle$ by constructing the corresponding matrix $\Gamma(z)$.
When computing the affinity, we must also include the contribution from the futile cycle fluxes. 
To calculate the futile cycle current $\langle J_{\rm fut}\rangle$, we construct the corresponding matrix $\Gamma^{\rm fut}(z)$, in which we multiply the non-diagonal entries corresponding to $k_{\text{PR},f}^{\text{C}}$, $k_{\text{PR},f}^{\text{NC}}$, $k_{\text{PR},r}^{\text{C}}$ and $k_{\text{PR},r}^{\text{NC}}$ by $e^z$, $e^z$, $e^{-z}$ and $e^{-z}$, respectively. 
Finally, the affinity of translation can be computed by Eq.~\ref{eqn:generalA}.

\subsection*{Stochastic simulation of DNA replication}
We simulated the replication of the first 300 base pairs of the T7 DNA polymerase gene sequence at the single molecule level, with Gillespie's algorithm \cite{Gillespie1977a}.
The simulation starts with the polymerase in the apo state at the 5' end of the gene sequence.
The only reactions available at this state are the binding reactions of the 4 dNTPs, which are assumed to be at equal concentrations.
After the binding of a dNTP, the simulation trajectories are generated based on the kinetic network shown in Fig.~\ref{fig:PolPlot}A.
After each polymerization reaction, the polymerase translocates on the DNA and reads the next nucleotide. 
The simulation is terminated when the $300^{\text{th}}$ nucleotide of the gene sequence is polymerized. 

The dynamics of DNA replication simulations are studied using the ensemble of trajectories generated. 
The total number of steps ($N_{rep}$) in completing the replication of the DNA sequence varies from one realization to another.
Selecting the completion time of replication ($\mathcal{T}$) as the output observable for each realization, we  define TUR of replication as
\begin{align}
\mathcal{Q}
= \left[ -\Delta\mu + \beta^{-1}I \right] \frac{ \langle (\delta \mathcal{T})^2 \rangle}{ \langle \mathcal{T} \rangle^2 },
\end{align}
where the dissipation has contributions from the free energy drive ($ \Delta\mu$) and Shannon-entropy ($I$). 
Denoting the forward and  reverse rate constants of each kinetic step by $k_{i,f}$ and $k_{i,r}$ for $i=1,\dots,N_{rep}$, we can compute the average free energy drive by
$-\beta \Delta\mu  =\langle\sum_{i=1}^{N_{rep}} \ln{\frac{k_{i,f}}{k_{i,r}}}\rangle$,
where $\langle\ldots\rangle$ denotes the average over the ensemble of $10^4$ realizations. 
The entropic contribution is computed as 
$I = -\sum_{l=1}^{300}\sum_{i_{\text{dNTP}}=1}^{4} \eta^l_{i_\text{dNTP}} \ln{\eta^l_{i_\text{dNTP}}}$,
where $\eta^l_{i_\text{dNTP}}$ is the probability of incorporating one of the 4 types of dNTPs, at the $l$-th position.

\subsection*{The ribosome model} 
We provide more details on the model of translation by the ribosome in Fig.~\ref{fig:Ribosome}. 
Our model is a modified version of that from Rudorf \emph{et. al.} \cite{Rudorf2014}, 
in which we combine all linear chains of consecutive and irreversible reactions into single reactions. 
For instance, consider two consecutive and irreversible reactions $(1) \rightarrow (2)$ and $(2) \rightarrow (3)$, with respective rate constants $k_{12}$ and $k_{23}$, defined among three states $(1)$, $(2)$ and $(3)$. If there are no other reactions associated with the state $(2)$, we remove the state $(2)$, and define a new reaction $(1) \rightarrow (3)$ with the rate constant $k_{13}^{-1} \equiv k_{12}^{-1}+ k_{23}^{-1}  $. 

To calculate the entropy productions, we defined reverse rate constants for all reactions, which were constrained by the affinity of the corresponding kinetic cycle. 
We assumed that affinities of the parallel kinetic cycles for the cognate and near-cognate aa-tRNAs were identical. 
The affinity of the futile cycle ($\Delta\mu_{\rm fut}$) comes from GTP hydrolysis, which dissipates $\approx 20$ $k_BT$ \cite{Berg2002}. Therefore, we applied the following constraints, where $-\beta\Delta\mu_{\rm fut}=20$.
\begin{align}
\ln{ \left( \frac{k_{\text{on}} [\text{C}] k_{ \text{rec},f}k^{\text{C}}_{ \text{hyd},f}k^{\text{C}}_{ \text{PR},f}}
{k_{\text{off}}k^{\text{C}}_{ \text{rec},r}k^{\text{C}}_{ \text{hyd},r}k^{\text{C}}_{{\text{PR}},r}[\text{C}']}\right) } &= -\beta\Delta\mu_{\rm fut}, 
\label{eqn:affinityconstraint1} \\
\ln{\left( \frac{k_{\text{on}} [\text{NC}] k_{ \text{rec},f}k^{\text{NC}}_{ \text{hyd},f}k^{\text{NC}}_{ \text{PR},f}}
{k_{\text{off}}k^{\text{NC}}_{ \text{rec},r}k^{\text{NC}}_{ \text{hyd},r}k^{\text{NC}}_{{\text{PR}},r}[\text{NC}']} \right)} &= -\beta\Delta\mu_{\rm fut}.
\label{eqn:affinityconstraint2} 
\end{align}

The affinity involved with polymerization ($\Delta\mu_{\text{pol}}$) can be estimated as the sum of the free energies of 
GTP hydrolysis, peptide bond formation, and the cleavage of the ester bond between the tRNA and the amino-acid.
The hydrolysis of the GTP molecule incurs the dissipation of $\approx 20$ $k_BT$. 
Conversely, each peptide bond synthesized stores $\approx 5$ $k_BT$ of free energy \cite{Martin1998}. 
%Although not shown in the kinetic network diagrams (Fig.~\ref{fig:RibosomeNetwork}), the formation of the aa-tRNA complex is driven by the the hydrolysis of ATP into AMP and inorganic pyrophosphate \cite{Berg2002}. 
The standard free energy of the ester bond between the amino-acid and the tRNA is $\approx 12$ $k_BT$ \cite{Loftfield1972,Berg2002}. 
Since the ratio of charged to uncharged tRNAs is $\approx 10$ fold \cite{Evans2017}, 
the net free energy of the ester bond between the amino-acid and tRNA is $\approx 15$ $k_BT$.
%Assuming that the free energy of the ester bond between the amino-acid and the tRNA is $15$ $k_BT$, 
Then, $-\beta\Delta\mu_{\text{pol}}\approx  30 $, which gives the following constraints
\begin{align}
\ln{ \left(\frac{k_{\text{on}}[\text{C}]k_{ \text{rec},f}k^{\text{C}}_{ \text{hyd},f}k^{\text{C}}_{\text{pol}}}{k_{\text{off}}k^{\text{C}}_{ \text{rec},r}k^{\text{C}}_{ \text{hyd},r}k^{\text{C}}_{\text{dep}}}\right) } &= -\beta\Delta\mu_{\text{pol}}, 
\label{eqn:affinityconstraint3} \\
\ln{ \left(\frac{k_{\text{on}}[\text{NC}]k_{ \text{rec},f}k^{\text{NC}}_{ \text{hyd},f}k^{\text{NC}}_{\text{pol}}}{k_{\text{off}}k^{\text{NC}}_{ \text{rec},r}k^{\text{NC}}_{ \text{hyd},r}k^{\text{NC}}_{\text{dep}}}\right) } &= -\beta\Delta\mu_{\text{pol}}.
\label{eqn:affinityconstraint4} 
\end{align}
The terms $k^{\rm C}_{\rm dep}$ and $k^{\rm NC}_{\rm dep}$ implicitly take into account the concentration of tRNA and (EF-Tu)-GDP, which detach during the final polymerization step.
In order to fully constrain all the rate constants, we set $k^{\text{C}}_{ \text{hyd},r} = 10^{-3} k^{\text{C}}_{ \text{hyd},f}$ and  $k^{\text{NC}}_{ \text{hyd},r} = 10^{-3} k^{\text{NC}}_{ \text{hyd},f}$. Modest changes to these affinity related constraints (Eqs~\ref{eqn:affinityconstraint1}-\ref{eqn:affinityconstraint4}) do not affect the qualitative conclusions of our work. 

\subsection*{Ternary complex concentration}

The ternary complex concentration was modeled as a function of the concentration of its components, 
aa-tRNA, EF-Tu, GTP, and GDP.
First, EF-Tu binds with GTP and GDP to form (EF-Tu)-GTP and (EF-Tu)-GDP, respectively. 
%Although this binding reaction is facilitated by another protein, EF-T, in cells, 
%since EF-T does not expend energy in this process, we will assume that EF-Tu, GTP, GDP, (EF-Tu)-GTP, and (EF-Tu)-GDP are present at equilibrium ratios. 
Then, aa-tRNA binds with (EF-Tu)-GTP and (EF-Tu)-GDP to form (aa-tRNA)-(EF-Tu)-GTP and (aa-tRNA)-(EF-Tu)-GDP, respectively \cite{Romero1985}.
Here, (aa-tRNA)-(EF-Tu)-GTP and (aa-tRNA)-(EF-Tu)-GDP represent the combined total of
all the respective 42 individual ternary complexes.
Assuming equilibrium,
%Assuming that all chemical species are present at their equilibrium ratios, 
we can write the following equalities,
\begin{align}
\rm [(aa\text{-}tRNA)\text{-}(EF\text{-}Tu)\text{-}GTP] &= \frac{\rm [aa\text{-}tRNA][(EF\text{-}Tu)\text{-}GTP]}{K_{\rm aaGTP}}, \label{eqn:GTP1}\\
\rm [(aa\text{-}tRNA)\text{-}(EF\text{-}Tu)\text{-}GDP] &= \frac{\rm [aa\text{-}tRNA][(EF\text{-}Tu)\text{-}GDP]}{K_{\rm aaGDP}}, \\
\rm [(EF\text{-}Tu)\text{-}(GTP)] &= \frac{\rm [EF\text{-}Tu][GTP]}{{K_{\rm GTP}}}, \\
\rm [(EF\text{-}Tu)\text{-}(GDP)] &= \frac{\rm [EF\text{-}Tu][GDP]}{{K_{\rm GDP}}} \label{eqn:GTP4},
\end{align}
where the respective dissociation constants are set to 
$  K_{\rm aaGTP} = {\rm 10^{-1} ~\mu M} ~ \text{\cite{Romero1985}},~
 K_{\rm aaGDP} = {\rm 14~ \mu M }~\text{\cite{Romero1985}},~
 K_{\rm GTP} = {\rm 6\times10^{-2}}~ \mu M~ \text{\cite{Gromadski2002}},~ {\rm and~}
 K_{\rm GDP} = {\rm  10^{-3} ~\mu M} ~ \text{\cite{Gromadski2002}}$. 
Using  Eqs.~\ref{eqn:GTP1}-\ref{eqn:GTP4}, we can solve for the concentration of all chemical species given the total [EF-Tu], [aa-tRNA], [GTP], and [GDP].
Unless specified otherwise, 
all the ribosome model plots (Figs.~\ref{fig:Ribosome} and \ref{fig:Translation}) 
are made at the cellular condition, with
$ \rm [EF\text{-}Tu]\rm=250~\mu M~ \text{\cite{Ishihama2008}},~
\rm [aa\text{-}tRNA] \rm \approx 200~\mu M~ \text{\cite{Dong1996}},~
\rm  [GTP]\rm =5000~\mu M ~\text{\cite{Bennett2009}},~and~
\rm [GDP]\rm = 700~\mu M~ \text{\cite{Bennett2009}}$.
At the cellular condition, the concentration of the GTP bound ternary complex is $\rm [(aa\text{-}tRNA)\text{-}(EF\text{-}Tu)\text{-}GTP] \approx [aa\text{-}tRNA] \approx 200~\mu\text{M}$. 

Assuming that all species of aa-tRNA were bound to (EF-Tu)-GTP and (EF-Tu)-GDP with equal binding constants $K_{\rm aaGTP}$ and $K_{\rm aaGDP}$,
we computed the concentration of individual ternary complexes by referencing the 
measured concentration ratios among individual aa-tRNA species.
For instance, the concentration of the cognate ternary complexes of codon UUU, which encodes for Phe, are
\begin{align}
 [\text{C}] &= \frac{ \rm [Phe\text{-}tRNA^{Phe}]_{\rm WT}}{ \rm [aa\text{-}tRNA]_{\rm WT} } [\rm(aa\text{-}tRNA)\text{-}(EF\text{-}Tu)\text{-}GTP],\\
 [\text{C}'] &= \frac{ \rm [Phe\text{-}tRNA^{Phe}]_{\rm WT}}{ \rm [aa\text{-}tRNA]_{\rm WT} } [\rm(aa\text{-}tRNA)\text{-}(EF\text{-}Tu)\text{-}GDP],
\end{align}
where the subscript WT represents the cellular concentrations obtained from Ref.~\cite{Dong1996}.

\begin{figure}[ht]
	\begin{center}
	\includegraphics[width=.5 \textwidth]{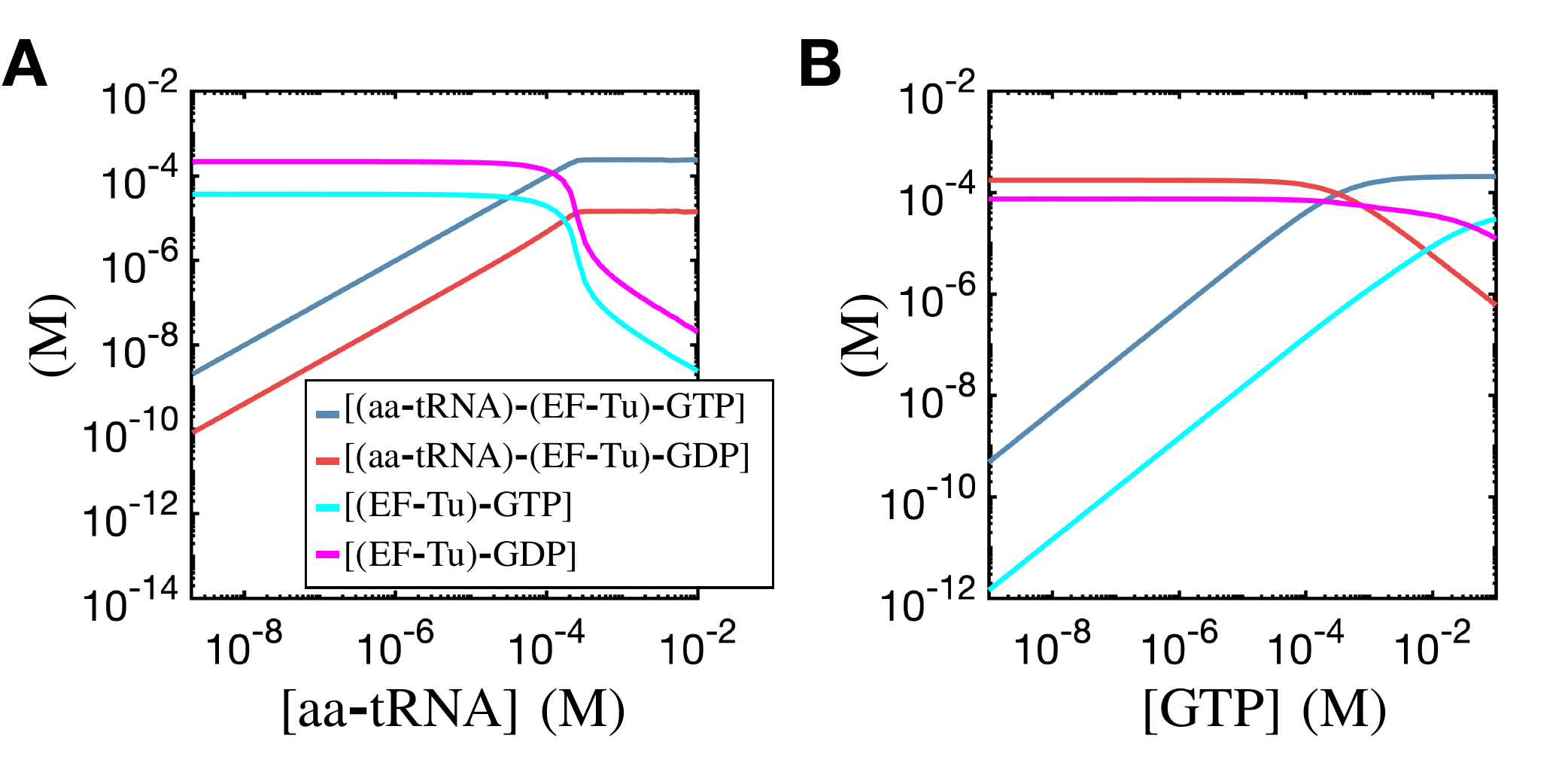}
	\caption{
	The concentration of the components of the ternary complex as functions of [aa-tRNA] and [GTP]. 
	(A) The dependence of [(aa-tRNA)-(EF-Tu)-GTP], [(aa-tRNA)-(EF-Tu)-GDP],  [(EF-Tu)-GTP], and  [(EF-Tu)-GDP] on [aa-tRNA],
	where [GTP] was fixed at 5 mM.
	(B) The dependence of [(aa-tRNA)-(EF-Tu)-GTP], [(aa-tRNA)-(EF-Tu)-GDP],  [(EF-Tu)-GTP], and  [(EF-Tu)-GDP] on [GTP],
	where [aa-tRNA] was fixed at $\approx0.2$mM. The curves are color-coded identically as those in (A).
	For both plots, we assumed cellular levels of [GDP](=0.7 mM) and [EF-TU](=0.25 mM),
	and used equilibrium dissociation constants as described in the SI text.
	}
	\label{fig:GTPdependence}
	\end{center}
\end{figure}

\subsection*{The extended model of translation}
Excluding the three stop codons, there are 61 types of codons encoding for 20 amino-acids.
For \textit{E. coli}, the 43 types of tRNAs were identified by Dong \emph{et. al} \cite{Dong1996}.
Out of these, the pairs Gly1-Gly2 and Ile1-Ile2 were not differentiated in the concentration measurements. 
In our simulations, we assumed that Gly1 and Gly2 (resp. Ile1 and Ile2) were each present in the cellular milieu at half of the measured concentration of the Gly1-Gly2 pair (resp. Ile1-Ile2 pair).
We removed the seleno-cysteine carrying tRNA from the analysis, since it is low in concentration, and it is incorporated into the polypeptide through a different kinetic scheme from the rest of the aa-tRNAs.
Overall, we included total 42 types of aa-tRNAs in the extended translation model,
with measurements from \textit{E. coli} dividing every $\approx$ 86 minutes \cite{Dong1996}.
The cognate, near-cognate, and non-cognate groupings of aa-tRNAs for each codon is shown in Fig.~\ref{fig:codon_table}.

mRNA translation by the ribosome at the single molecule level is simulated with Gillespie's algorithm \cite{Gillespie1977a}.
The simulation starts with the ribosome in the apo state at the start codon. 
The only reactions available at this state are the bindings of the 42 aa-tRNAs, the concentrations of which were taken from Dong \emph{et. al.} \cite{Dong1996}. 
After the binding of an aa-tRNA, the simulation 
trajectories were generated on the kinetic network
shown in Fig.~\ref{fig:Ribosome}B.
After each polymerization reaction, the ribosome reads the next codon, translocating along the mRNA. 
The simulation is terminated when the ribosome completes the translation of the last codon. 

\subsection*{The Hopfield model}
\begin{figure}[ht]
	\begin{center}
	\includegraphics[width=.5 \textwidth]{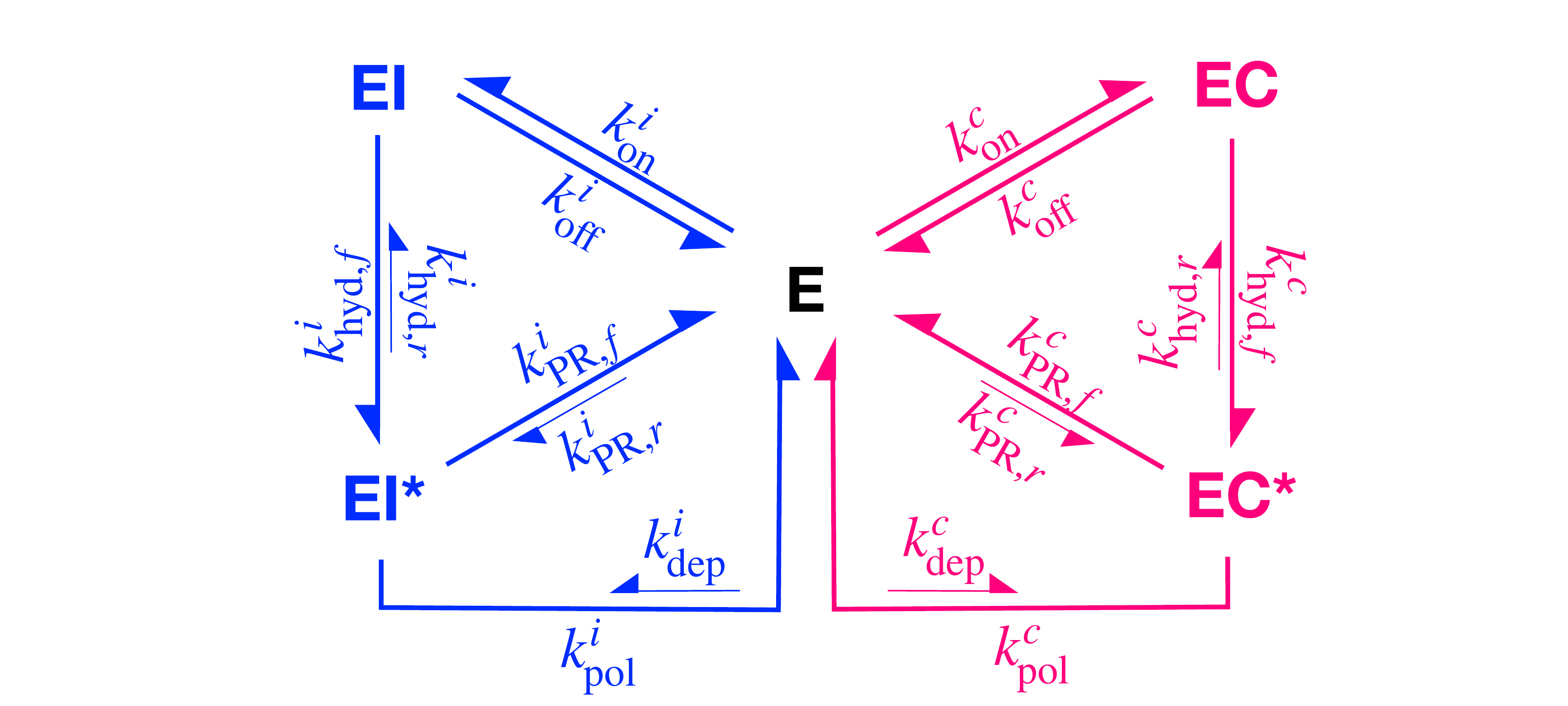}
	\caption{
	Schematic of the modified Hopfield model \cite{Hopfield1974}.}
	\label{fig:HopfieldSchematic}
	\end{center}
\end{figure}

We provide more details on the modified Hopfield model \cite{Hopfield1974}.
The reaction cycle of the Hopfield model is composed of substrate binding (${\bf E}+{\bf C}\rightleftharpoons {\bf EC}$ and ${\bf E}+{\bf I}\rightleftharpoons {\bf EI}$), 
followed by the effectively irreversible steps of ATP hydrolysis (${\bf EC }\rightleftharpoons{\bf EC^*}$ and ${\bf EI }\rightleftharpoons{\bf EI^*}$)
and polymerization (${\bf EC^*}\rightleftharpoons{\bf E}$ and ${\bf EI^*}\rightleftharpoons{\bf E}$) (Fig.~\ref{fig:HopfieldSchematic}). 
At states ${\bf EC^*}$ and ${\bf EI^*}$, the substrate can also dissociate through the proofreading reaction (PR, ${\bf EC^*}\rightleftharpoons{\bf E}+{\bf C}$ and ${\bf EI^*}\rightleftharpoons{\bf E}+{\bf I}$), which is also effectively irreversible. 

\begin{figure*}[ht]
\begin{center}
	\includegraphics[width=.95 \textwidth]{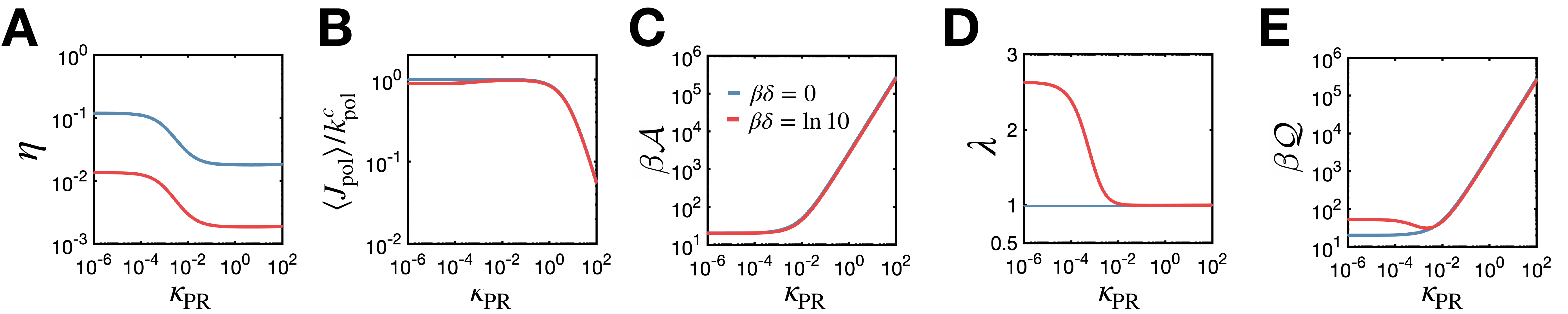}
	\caption{ 
	Dynamical properties of the modified Hopfield model with kinetic discrimination. 
	(A) The error probability ($\eta$), (B) normalized polymerization current ($\langle J_{\text{pol}}\rangle /k^c_{\text{pol}}$),
	(C) affinity ($\mathcal{A}$), (D) Fano factor ($\lambda$), and (E) $\mathcal{Q}$ as functions of $\kappa_{\text{PR}}$.
	For (A)-(E) The perturbative parameter $\kappa_{\text{PR}}$ is multiplied to the rates $k^c_{ {\text{PR}},f}$, $k^i_{ {\text{PR}},f}$, $k^c_{ {\text{PR}},r}$ and $k^i_{ {\text{PR}},r}$.
	The blue line presents the original Hopfield model with $\beta\delta=0$, 
	and the orange line represents the modified Hopfield model with kinetic discrimination, with $\beta\delta=\ln10$. 
	In (C), the orange and blue lines are nearly identical.
	The rate constants used to make these plots are given in Table S3. 
	}
	\label{fig:HopfieldSI}
	\end{center}
\end{figure*}

In the original formulation of the Hopfield model, the forward kinetic rate constants of correct and incorrect pathways were identical.
In our modified version, the forward constants satisfy the following relations
\begin{align}
e^{\beta\delta}=\frac{k^c_{\text{on} }}{k^i_{\text{on}}} = \frac{k^c_{ {\text{hyd}},f }}{k^i_{{\text{hyd}},f}} =\frac{k^c_{\text{pol} }}{k^i_{\text{pol}}}= \frac{k^c_{ {\text{PR}},r }}{k^i_{ {\text{PR}},r }},
\end{align}
with $\beta\delta>0$. 
Next, to allow for error reduction by proofreading, 
we set the forward kinetic rates so that $k^c_{\text{pol}}\ll k^c_{\text{hyd},f} \ll k^c_{\text{on}}$.
Finally, we constrained the reverse reaction rates so that the affinities associated with the kinetic cycles are 
$\Delta\mu^{c}_{\text{pol}}= -20~k_BT$, and $\Delta\mu^i_{\text{pol}} = \Delta\mu^c_{\rm fut} = \Delta\mu^i_{\rm fut}= -18~k_BT$. 
The rate constants used to generate Fig.~\ref{fig:Hopfield} and Fig.~\ref{fig:HopfieldSI} are provided in Table S3.  

% Figures 

\begin{figure*}[ht]
	\begin{center}
	\includegraphics[width=0.95 \textwidth]{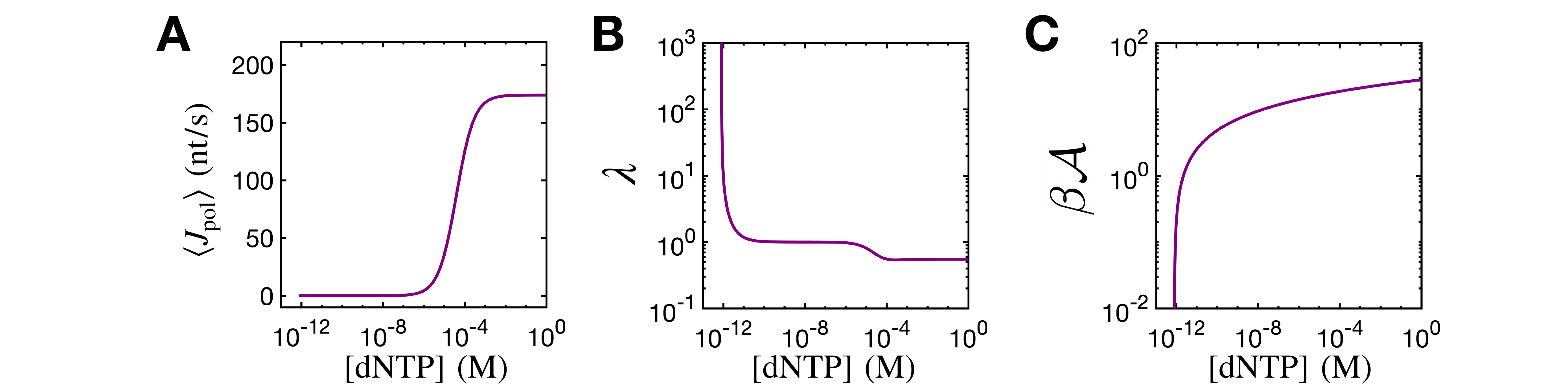}
	\caption{
	Dynamical properties of the exonuclease-deficient T7 DNA polymerase, obtained at identical conditions as in Fig.~\ref{fig:PolPlot}. 
	{(A)} The reaction current ($\langle J_{\text{pol}} \rangle$), (B) Fano factor of the reaction current ($\lambda$), and (C) the affinity ($\mathcal{A}$) are plotted against [dNTP]. }
	\label{fig:PolSup}
	\end{center}
\end{figure*}

\begin{figure*}[ht]
	\begin{center}
	\includegraphics[width=.95 \textwidth]{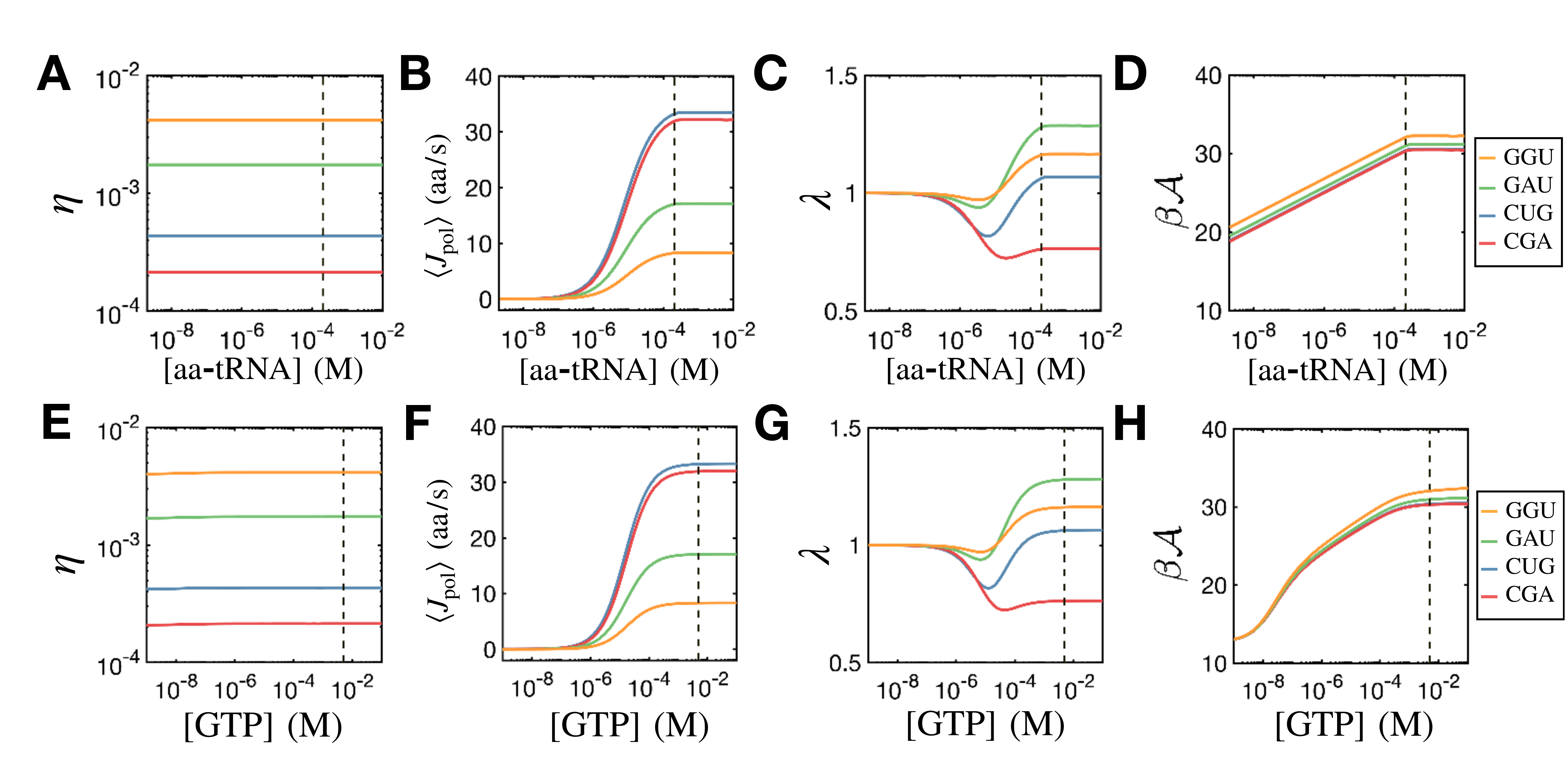}
	\caption{
	Dynamical properties of mRNA translation by \textit{E. coli} ribosome.
	(A) The error probability ($\eta$), (B) polymerization current $(\langle J_{\text{pol}} \rangle)$, (C) Fano factor of the polymerization current ($\lambda$),
	and (D) affinity ($\mathcal{A}$) plotted against [aa-tRNA], at identical conditions as in Fig.~\ref{fig:Ribosome}D. 
	(E) The error probability ($\eta$), (F) polymerization current $(\langle J_{\text{pol}} \rangle)$, (G) Fano factor of the polymerization current ($\lambda$),
	and (H) affinity ($\mathcal{A}$) plotted against [GTP], at identical conditions as in Fig.~\ref{fig:Ribosome}E. 
	For all plots, the dashed black line represents the cellular condition in \textit{E. coli}.
	}
	\label{fig:RibosomeSup}
	\end{center}
\end{figure*}

\begin{figure*}[ht]
	\begin{center}
	\includegraphics[width=.9 \textwidth]{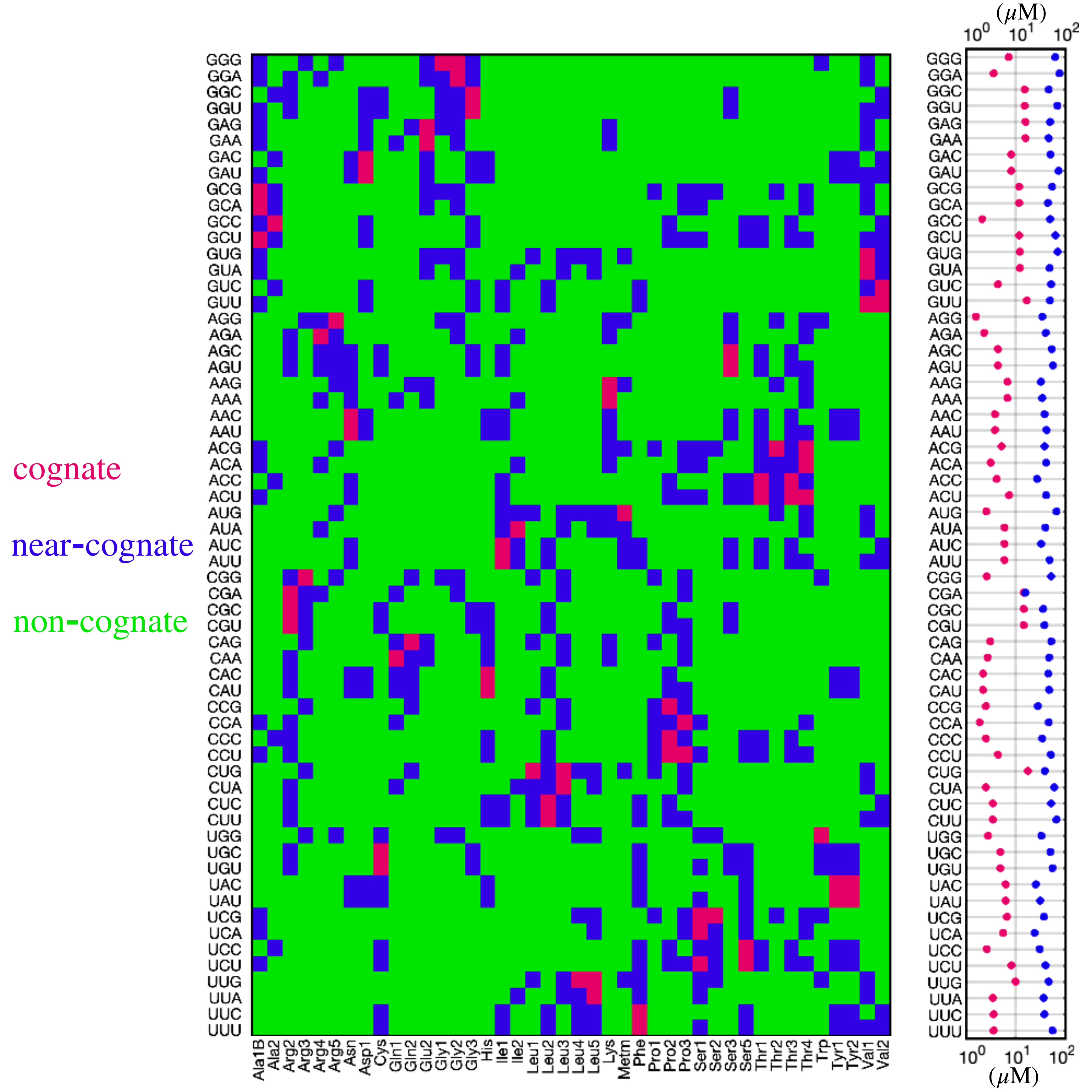}
	\caption{
	(Left) The groupings of cognate (red), near-cognate (blue), and non-cognate (green) aa-tRNAs for each codon \cite{Rudorf2014}.
	(Right) The sum of the concentration of the cognate (red) and near-cognate (blue) aa-tRNA species, plotted for each codon  \cite{Dong1996}.
	}
	\label{fig:codon_table}
	\end{center}
\end{figure*}

\begin{figure*}[ht]
	\begin{center}
	\includegraphics[width=.95 \textwidth]{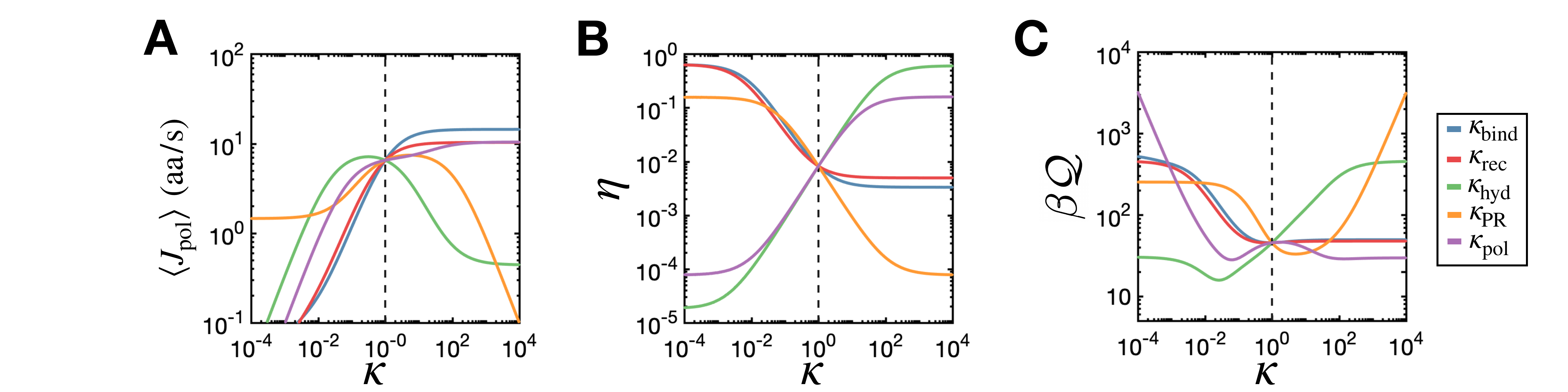}
	\caption{
	Dynamical properties of mRNA translation by the $\textit{E. coli}$ ribosome with respect to perturbations to the wild type rate constants.
	(A) The polymerization current ($\langle J_{\text{pol}} \rangle$), (B) the error probability ($\eta$), and (C) $\mathcal{Q}$ as functions of $\kappa_{ \rm bind}$, $\kappa_{\text{rec}}$, $\kappa_{\text{hyd}}$, $\kappa_{\text{PR}}$, and $\kappa_{\text{pol}}$.
	The perturbative parameters $\kappa_{ \rm bind}$, $\kappa_{\text{rec}}$, $\kappa_{\text{hyd}}$, $\kappa_{\text{PR}}$, and $\kappa_{\text{pol}}$ were each multiplied to the reactions associated with binding, codon-recognition, GTP-hydrolysis, proofreading, and polymerization, respectively. 
	The black dashed lines represent the wild type condition for the codon UUU.
	}
	\label{fig:RibosomeSup3}
	\end{center}
\end{figure*}

\clearpage

% Tables
\begin{table}[ht]
\begin{center}
\caption{Rate constants for the T7 DNA polymerase, from Ref. \cite{Tsai2006}. The rate constants $ k^c_{\text{dep}}$ and $ k^i_{\text{dep}}$ are determined such that the affinities of correct and incorrect monomer incorporations are $20$ and $15$ $k_BT$, respectively, when [dNTP]=$100$ $\mu$M \cite{Goodman1997,Minetti2003PNAS}.
The terms $k^{c}_{\rm dep}$ and $k^{i}_{\rm dep}$ implicitly take into account the concentration of PPi which detaches during the final polymerization step.
}
\begin{tabular}{c | c} 
 \hline
$ k^c_{\text{on}}$ & $10^2$ $ {\rm \mu {M}^{-1}}s^{-1}$ \\  
\hline
$ k^c_{ {\text{conf}},f}$ & $6.0 \times 10^2 $ $s^{-1}$ \\  
 \hline
$ k^c_{\text{pol}}$ & $3.6 \times 10^2 $ $s^{-1}$ \\  
 \hline
$ k^c_{\text{off}}$ & $2.8 \times 10^3 $ $s^{-1}$ \\  
 \hline
$ k^c_{ {\text{conf}},r}$ & $1.6$ $s^{-1}$ \\  
 \hline
$ k^c_{\text{dep}}$ & $10^{-3}$ $s^{-1}$ \\  
 \hline
$ k^i_{\text{on}}$ & $10^2$ ${\rm \mu {M}^{-1}}s^{-1}$ \\  
\hline
$ k^i_{ {\text{conf}},f}$ & $2.2 \times 10^2 $ $s^{-1}$ \\  
 \hline
$ k^i_{\text{pol}}$ & $3.0 \times 10^{-1} $ $s^{-1}$ \\  
 \hline
$ k^i_{\text{off}}$ & $2.0 \times 10^4 $ $s^{-1}$ \\  
 \hline
$ k^i_{ {\text{conf}},r}$ & $4.2 \times 10^2 $ $s^{-1}$ \\  
 \hline
$ k^i_{\text{dep}}$ & $2.4\times10^{-8}$ $s^{-1}$ \\  
 \hline
\end{tabular}
\end{center}
\end{table}

\begin{table}[ht]
\begin{center}
\caption{Rate constants for the wild type \textit{E. coli} ribosome, from Ref. \cite{Rudorf2014}. 
The rate constants
$ k^{\text{C}}_{ {\text{PR}},r}$, ${k}^{\text{C}}_{\text{dep}}$, $ k^{\text{NC}}_{{\text{PR}},r}$, and ${k}^{\text{NC}}_{\text{dep}}$
were determined from the constraints associated with the affinity of the kinetic cycles at the wild type condition (Eqs~\ref{eqn:affinityconstraint1}-\ref{eqn:affinityconstraint4}).}
\begin{tabular}{c | c} 
 \hline
$ k_{\text{on}}$ & $9.4\times10$ $\mu \text{M}^{-1}s^{-1}$ \\  
\hline
$ k_{\text{off}}$ & $1.4\times10^3$ $s^{-1}$ \\  
 \hline
$ {k}_{ {\text{rec}}, f}$ & $2.1 \times 10^{3} $ $s^{-1}$ \\  
 \hline
$ {k}^{\text{C}}_{ {\text{rec}},r}$ & $2$ $s^{-1}$ \\  
 \hline
$ {k}^{\text{C}}_{ {\text{hyd}},f}$ & $3.75\times10^2$ $s^{-1}$ \\  
 \hline
$ {k}^{\text{C}}_{ {\text{hyd}},r}$ & $k^{\text{C}}_{ {\text{hyd}},f}\times10^{-3}$\\  
 \hline
$ k^{\text{C}}_{ {\text{PR}},f}$ & $1$ $s^{-1}$ \\  
 \hline
${k}^{\text{C}}_{\text{pol}}$ & $1.1\times10^{2}$ $s^{-1}$ \\  
 \hline
$ {k}^{\text{NC}}_{ {\text{rec}},r}$ & $2.7\times 10^{3} $ $s^{-1}$ \\  
 \hline
$ {k}^{\text{NC}}_{{\text{hyd}},f}$ & $4.9$ $s^{-1}$ \\  
 \hline
$ {k}^{\text{NC}}_{{\text{hyd}},r}$ & $k^{\text{NC}}_{ {\text{hyd}},f} \times10^{-3}$\\  
 \hline
$ k^{\text{NC}}_{{\text{PR}},f}$ & $6$ $s^{-1}$ \\  
 \hline
${k}^{\text{NC}}_{\text{pol}}$ & $2.7\times10^{-1}$ $s^{-1}$ \\   [1ex] 
 \hline
 \end{tabular}
\end{center}
\end{table}

\begin{table}[ht]
\begin{center}
\caption{Rate constants for the Hopfield model with kinetic discrimination, 
where $-\beta\Delta\mu_c=2$ and $-\beta\Delta\mu_i=0$.
The parameters $\delta$ and $\kappa_{\text{PR}}$ are as defined in the main text.}
\begin{tabular}{c | c} 
\hline
$ k^c_{\text{on}}$ & $10^{3}~s^{-1}$ \\  
\hline
$ k^c_{\text{off}}$ & $k^c_{\text{on}}e^{\beta\Delta\mu_c}$ \\  
\hline
$ k^c_{\text{hyd},f}$ & $1~s^{-1}$ \\  
\hline
$ k^c_{\text{hyd},r}$ & $k^c_{\text{hyd},f}e^{-10}$ \\  
\hline
$ k^c_{\text{pol}}$ & $10^{-3}~s^{-1}$ \\  
\hline
$ k^c_{\text{dep}}$ & $k^c_{\text{pol}}e^{-8}$ \\  
\hline
$ k^c_{\text{PR},f}$ & $  \kappa_{\text{PR}} e^{\beta\Delta\mu_c}~s^{-1}$ \\  
\hline
$ k^c_{\text{PR},r}$ & $ k^c_{\text{PR},f} e^{-8 - \beta\Delta\mu_c} $ \\  
\hline
$ k^i_{\text{on}}$ & $k^c_{\text{on}}e^{-\beta\delta} $ \\  
\hline
$ k^i_{\text{off}}$ & $k^i_{\text{on}}e^{\beta\Delta\mu_i}$ \\  
\hline
$ k^i_{\text{hyd},f}$ & $k^c_{\text{hyd},f}e^{-\beta\delta} $ \\  
\hline
$ k^i_{\text{hyd},r}$ & $k^i_{\text{hyd},f}e^{-10}$ \\  
\hline
$ k^i_{\text{pol}}$ & $k^c_{\text{pol}}e^{-\beta\delta}  $ \\  
\hline
$ k^i_{\text{dep}}$ & $k^i_{\text{pol}}e^{-8}$ \\  
\hline
$ k^i_{\text{PR},f}$ & $\kappa_{\text{PR}} e^{-\beta\delta +\beta\Delta\mu_i}~s^{-1}$ \\  
\hline
$ k^i_{\text{PR},r}$ & $ k^i_{\text{PR},f}e^{-8-\beta\Delta\mu_i}~s^{-1}$ \\     [1ex] 
 \hline
 \end{tabular}
\end{center}
\end{table}

%\clearpage
%\bibliography{library,mybib1}

\end{document}